\documentclass[numberedappendix,twocolappendix,apj,iop]{emulateapj}
\usepackage{amsmath}
\usepackage{bm}

\newcommand{\new}{}

\shorttitle{DYNAMICS OF CO IN AMORPHOUS WATER ICE ENVIRONMENTS}
\shortauthors{Karssemeijer et al.}

\begin{document}

\title{DYNAMICS OF CO IN AMORPHOUS WATER ICE ENVIRONMENTS}

\author{L.J. Karssemeijer\altaffilmark{1}, S. Ioppolo\altaffilmark{1,2}, M.C. van Hemert\altaffilmark{3}, A. van der Avoird\altaffilmark{1},  M.A. Allodi\altaffilmark{4}, G.A. Blake\altaffilmark{2,4} and H.M. Cuppen\altaffilmark{1,5} }


\altaffiltext{1}{Theoretical Chemistry, Institute for Molecules and Materials, Radboud University Nijmegen, Heyendaalseweg 135, 6525 AJ Nijmegen, The Netherlands.}
\altaffiltext{2}{Division of Geology and Planetary Science, California Institute of Technology, 1200 E. California Blvd., Pasadena, CA 91125, USA}
\altaffiltext{3}{Gorlaeus Laboratories, Leiden Institute of Chemistry, Leiden University, P.O. Box 9502, 2300 RA Leiden, The Netherlands}
\altaffiltext{4}{Division of Chemistry and Chemical Engineering, California Institute of Technology, 1200 E. California Blvd., Pasadena, CA 91125, USA}
\altaffiltext{5}{h.cuppen@science.ru.nl}


\begin{abstract}
The long-timescale behavior of adsorbed carbon monoxide on the surface of amorphous water ice is studied under dense cloud conditions by means of off-lattice, on-the-fly, kinetic Monte Carlo simulations. It is found that the CO mobility is strongly influenced by the morphology of the ice substrate. Nanopores on the surface provide strong binding sites which can effectively immobilize the adsorbates at low coverage. As the coverage increases, these strong binding sites are gradually occupied leaving a number of admolecules with the ability to diffuse over the surface. Binding energies, and the energy barrier for diffusion are extracted for various coverages. Additionally, the mobility of CO is determined from isothermal desorption experiments. Reasonable agreement on the diffusivity of CO is found with the simulations. Analysis of the 2152~cm$^{-1}$, polar CO band supports the computational findings that the pores in the water ice provide the strongest binding sites and dominate diffusion at low temperatures.
\end{abstract}


\section{INTRODUCTION}
Molecules are important constituents of the interstellar medium (ISM). In molecular clouds, they function as coolants and can be used to trace physical conditions like temperature, hydrogen density, and the lifetime of the cloud~\citep{Herbst2009a}. Furthermore, simple interstellar molecules are necessary precursors to more complex biomolecules which are essential for the formation of life~\citep{Charnley1992,VanDishoeck1998}. 

In the ISM, a rich gas phase chemistry leads predominantly to unsaturated (organic) molecules whereas saturated molecules are mostly formed on dust grain surfaces. The diffusive Langmuir-Hinshelwood mechanism is the primary formation process on ice mantles. As such, the diffusion of reactants is of key importance~\citep{Hagasewa1992,Garrod:2008b} and should be well understood. In many rate equation models, the rate of a simple addition reaction $\textrm{A}+\textrm{B}\to \textrm{C}$ is implemented as
\begin{equation}
\frac{\mathrm{d}\left[\textrm{C}\right]}{\mathrm{d}t} = k_{\textrm{\scriptsize{A+B}}}\left(k_{\textrm{\scriptsize{D,A}}}+k_{\textrm{\scriptsize{D,B}}}\right)\left[\textrm{A}\right]\left[\textrm{B}\right], \label{Dre}
\end{equation}
where the square brackets denote the surface concentrations of species. For reactions with a barrier, $k_{\textrm{\scriptsize{A+B}}}$ is the rate for crossing this barrier and $k_{\textrm{\scriptsize{D,A}}}$ specifies the rate with which species $\textrm{A}$ scans the grain surface. Eq~\eqref{Dre} demonstrates the importance of accurate diffusion rates. Experimentally however, measurements of diffusion rates are tedious and often prone to errors. This has led to a lack of accurate information, leaving diffusion as an uncertain process in many models. Diffusion barriers are now frequently assumed to be a fixed fraction of the surface binding energy. Even though this fraction influences outcomes of the models significantly~\citep{Vasyunin2013}, a wide range of values is used, ranging from 0.3 to 0.8~\citep{Hagasewa1992,Ruffle2002,Cuppen:2009,Chang2012}. Theoretical studies into the dynamical behavior of specific adsorbates on model interstellar ices can thus greatly add to the value of rate equation models by specifically calculating input parameters like energy barriers for diffusion and desorption. 

In previous work, we have studied the mobility of CO on hexagonal water ice and found that, for this system, the diffusion barrier is equal to one third of the binding energy~\citep{Karssemeijer2012}. In the present paper we will apply the same methodology to determine the diffusion barriers of a more astrophysically relevant system: adsorbed CO on amorphous solid water (ASW). Furthermore, we will present results from isothermal desorption experiments, from which we determined the diffusion coefficient of CO in amorphous water ice environments.

There are several reasons for taking H$_2$O\sbond CO as a model system. Firstly, CO is the second most abundant molecule in the ISM and is a key precursor for more complicated species like carbon dioxide~\citep{Ioppolo2011}, methanol~\citep{Watanabe2002}, and formic acid~\citep{Ioppolo2011a}. All these reactions occur on ice mantles of which H$_2$O is the main component~\citep{Gibb2000} so the dynamics of adsorbed CO on water ices forms an integral part of the molecular cloud's chemistry. Secondly, because CO forms in the gas phase and freezes out on the grain mantles only under certain conditions~\citep{Pontoppidan2006}, the ice mantles probably have a layered structure~\citep{Allamandola1999,Cuppen2011B,Oberg2011} where the interface between the layers may be especially interesting for astrochemistry. Also, given its importance in the chemical evolution of molecular clouds, the  H$_2$O\sbond CO system has been extensively studied both experimentally~\citep{Bar-nun1985,Devlin1992,Allouche1998,Manca2000,Collings2003b,Ayotte2001} and theoretically~\citep{Al-Halabi2004a,Al-Halabi2004,Manca2001} so there is ample reference material.

Finally, we can compare our results on amorphous systems to our previous study on ice Ih to learn about the importance of the surface structure of the ice substrate. This difference between crystalline and amorphous substrates is interesting because this is not examined in astrochemical models, which assume homogeneous grain surfaces, nor in laboratory experiments, which only probe average properties. Also given the amount of discussion about the porosity of interstellar ices~\citep{Bossa2012,Ayotte2001}, we will emphasize the effect of surface inhomogeneity on the diffusion coefficient and binding energy of adsorbed CO.

The first experimental measurements of the diffusion coefficient of CO in ASW were reported only very recently by \cite{Mispelaer2013}, as part of a larger set of studied species. In our experiments, we use a similar technique but we have focused only on CO and studied a broader temperature window to get a more accurate value for the diffusion barrier. Theoretically, no simulations reaching beyond the molecular dynamics timescale (roughly nanoseconds) have yet been performed on amorphous substrates. We will present these simulations using an off-lattice kinetic Monte Carlo (KMC) approach which can probe long timescales but which also has a high level of detail because it gives access to the positions of all individual atoms throughout the simulation. We will show that the surface structure, and especially its porosity, plays a critical role in the CO mobility making the amorphous system essentially different from crystalline systems we studied before. 

We will start in Section~\ref{sec:simulations} with a description of the KMC simulations and their results. In Section~\ref{sec:experiments}, the experiments are presented and these are compared with the outcome of the simulations in Section~\ref{sec:comparison}. Astrophysical implications,  as well as the consequences for larger scale astrochemical models are discussed Section \ref{sec:astrophysical}. The conclusions are summarized in Section~\ref{sec:conclusions}.

\section{SIMULATIONS}
\label{sec:simulations}
The dynamics of the water-CO systems are simulated with the Adaptive Kinetic Monte Carlo (AKMC) technique~\citep{Henkelman2001,Pedersen2010}, which combines the atomistic level of detail from molecular dynamics with the ability of probing long timescales from KMC simulations. The technique is implemented in the EON code\footnote{\url{http://theory.cm.utexas.edu/eon/
}} and is described in more detail in \cite{Karssemeijer2012}, which also demonstrates its applicability to molecular systems. In this section we will only give a short summary of the AKMC method before proceeding to the simulations themselves, and their results. 

\subsection{Computational Details}
\label{sec:compdetails}
The time evolution of a system in AKMC is described in exactly the same way as in normal KMC simulations~\citep{Bortz1975,Gillespie1976,Charnley1998,Chang2005}. A sequence of steps between discrete states with corresponding time increments is generated based on computer-generated, pseudorandom numbers. This procedure requires every state to have a unique table of events (TOE). This table contains all possible processes and their rates, which can take the system out of its current state, into the next. In traditional KMC, states are defined by a set of occupation numbers on a lattice and the TOEs have to be specified before the start of the simulation. In AKMC however, this is not the case. States are defined as local minima on a potential energy surface (PES) which in turn\new{,} relies on atomic coordinates through any kind of force field or higher level method. Hence, the particles are not confined to lattice positions and atomistic details of the system are available (the positions of all individual atoms are known). New states are discovered on-the-fly, as the simulation progresses, by performing swarms of transition-state searches on the PES which fill the TOEs by calculating transition rates. In this work\new{,} the transition states are first-order saddle points (SPs) on the PES. The minimum-mode following method~\citep{Henkelman1999,Malek2000,Olsen2004} is used to locate the SPs and rates are estimated using harmonic transition state theory (HTST)~\citep{Vineyard1957}.

\new{As explained in our previous paper~\citep{Karssemeijer2012}, an advantage of the KMC method is that, once the TOEs of a system are known for a specific temperature, one can easily adjust them for simulations at different temperatures, without having to perform new transition-state searches. We therefore typically perform AKMC simulations of a system at a high temperature first, where states are easily discovered. We then use the resulting TOEs for KMC simulations at lower temperatures. From the simulations, we obtain the trajectories of adsorbed CO molecules on ASW substrates. The mean squared displacements of these trajectories are then used to extract the diffusion coefficient of the adsorbates.}

Interactions in the system are described by means of classical pair potentials for both inter- and intramolecular forces. Since we have two molecular species in our systems, three interaction potentials are needed: H$_2$O\sbond H$_2$O, CO\sbond CO, and H$_2$O\sbond CO. The details of all three potentials are given in Appendix~\ref{app:cocopot}. Because the H$_2$O\sbond CO and CO\sbond CO potentials were fitted directly to \textit{ab-initio} calculations, corrections were made to account for the zero point energy contribution for all binding energy calculations in this work. This procedure is outlined in Appendix~\ref{app:zpe}.

In this study, three different amorphous ice substrates were studied. \new{Each of these has a unique surface morphology, due to their different initial structures. By using three different substrates, instead of one, we get a better handle on the spread in the results due to a particular surface morphology.} The substrates were prepared in the following manner using molecular dynamics simulations. An initial sample is generated containing 480 water molecules with a density of 0.94 $\textrm{g cm}^{-3}$, the experimental density of low-temperature vapor-deposited H$_2$O ice~\citep{Jenniskens1994}, at random (non-overlapping) positions with orientations such that the entire system has no net dipole moment. Periodic boundary conditions are applied along all three Cartesian axes. This system is equilibrated at 300~K using a Nos\'e-Hoover thermostat for 100 ps after which the thermostat is instantaneously set to 10~K and the system is left for another 20~ps. The thermostat is then turned off and the system is left to equilibrate for another 100~ps after which the bottom 120 molecules are constrained to their \new{instantaneous} positions in this bulk amorphous structure. Next, the periodic boundary condition in the $z$-direction is removed to create a surface and the temperature is increased to 100~K over a period of 100~ps. \new{In the $x$- and $y$-directions, the periodic boundary conditions remain applied in order to mimic an infinitely large surface. After this}, the system is once again equilibrated for 100~ps, then cooled back to 10~K in 100~ps\new{,} and left to equilibrate for another 100~ps. \new{All heating, cooling, and equilibration periods in this procedure were chosen sufficiently long to stabilize the energy fluctuations in the system to their expected values.} Finally the system is relaxed to the nearest minimum on the PES. The three different amorphous ice substrates generated in this way will be referred to as $S_1$, $S_2$, and $S_3$. 

Even though the substrates each contain 360 water molecules which are free to move, we will constrain their coordinates in some of the simulations. In this case, an additional superscript $c$ will be added to the substrate name.

\subsection{Computational Results}
The discussion of the simulation results starts with an investigation of characteristics of the amorphous ice substrates and the nature of the CO binding sites on their surfaces. We then discuss the dynamics of a single adsorbed CO molecule on each surface and evaluate the effect of the CO surface coverage. These results will be directly relevant for surface chemistry in the ISM and can be compared to the diffusion measurements. Finally, we turn our attention to systems with multiple adsorbed CO molecules, corresponding to the late stage in cloud evolution, where CO-dominated ice layers start to form~\citep{Oberg2011}. These systems allow for comparison to TPD experiments reported in literature.

\begin{figure*}[t!]
\includegraphics{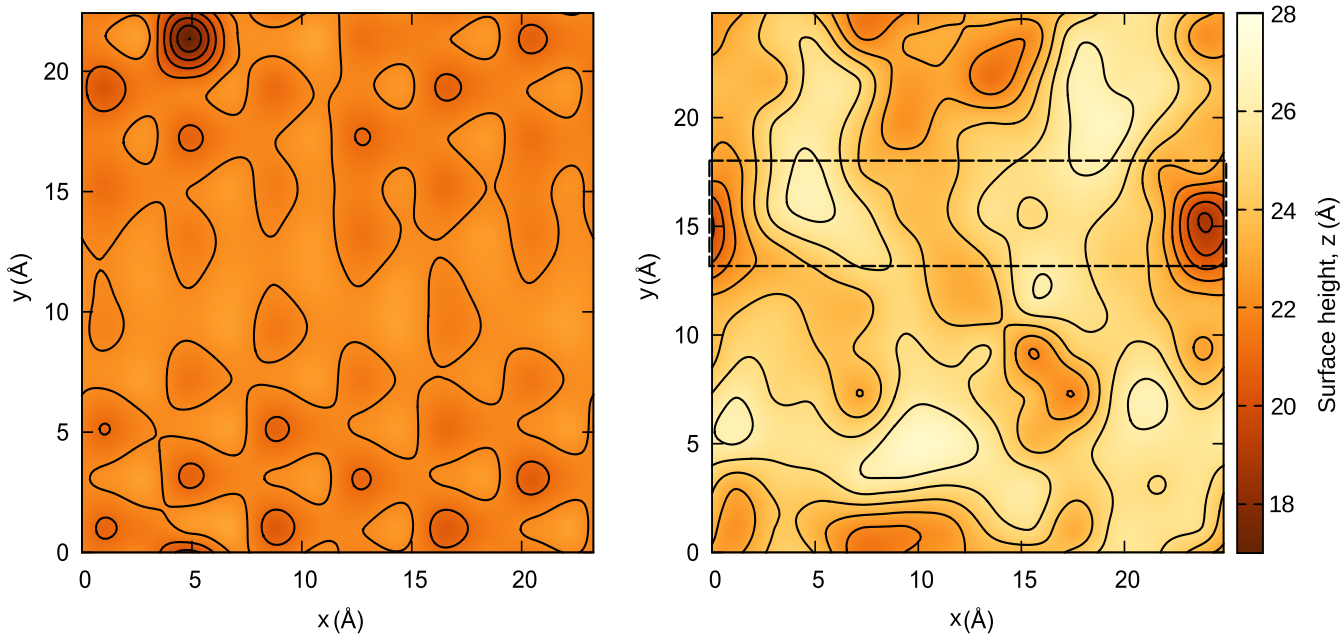}
\caption{Surface height map of ice Ih sample 1 from \cite{Karssemeijer2012} (left) and amorphous substrate $S_1$ (right). The dashed contour on the amorphous substrate shows the cross section from Figure~\ref{fig:co_pore}. The $z$-coordinate is measured from the lowest atomic coordinate in the system. \new{Periodic boundary conditions are applied in the $x$- and $y$-directions.}\label{fig:height_surface}}
\end{figure*}

\begin{figure}[h]
\includegraphics{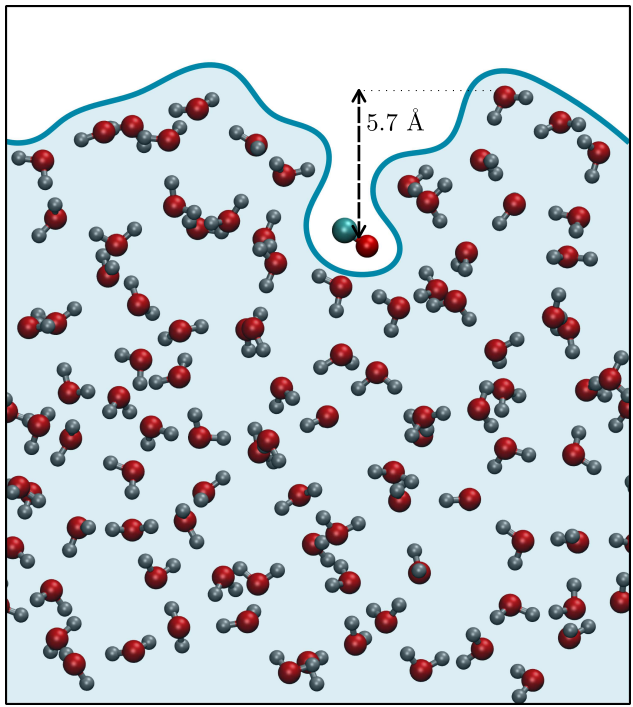}
\caption{Typical cross section of an amorphous ice substrate with a surface nanopore containing a CO molecule. The figure corresponds to an actual state found from AKMC simulations on substrate $S_1$. The cross section is also indicated in Figure~\ref{fig:height_surface}, where the surface pore is clearly visible (although it has been shifted for clarity)\label{fig:co_pore}.}
\end{figure}

\subsubsection{Substrate Morphology}
\label{sec:morphology}
The morphology of the ASW substrates is expected to determine the behavior of any adsorbate. We therefore start with an analysis of the bulk density and porosity of the substrates. The bulk density of the ice was obtained by averaging the density in a set of spheres of radius 4~\AA, centered on a regular grid with a lattice spacing of about 2~\AA. With increasing $z$-coordinate, the density in the spheres drops because they approach the rough, nanoporous surface. If we leave these spheres out of the calculations, to avoid effects of the surface, we find a bulk density value of $1.01\pm0.03$~g~cm$^{-3}$. There is little variation between the three different substrates. The density is somewhat higher than the experimental value for vapor deposited ice of 0.94~g~cm$^{-3}$~\citep{Jenniskens1994}. This is probably explained by the presence of bulk macropores in the laboratory ices, which are absent in our samples. 

The porosity of the surface itself is illustrated in Figure~\ref{fig:height_surface}. This figure shows the height of the surface for the amorphous sample $S_1$, as well as for a hexagonal ice sample (containing 360 water molecules). The surface height was calculated from a regular grid in the $x,y$-plane with a spacing of about 2~\AA. Each point on this grid was then assigned a $z$-value,\new{ the surface height,} corresponding to the point 2.5~\AA\ (roughly the H$_2$O\sbond CO distance) away from the center of mass of the nearest substrate molecule. The amorphous sample clearly has a much larger surface roughness than the crystalline sample. Its $z$-coordinate covers an almost 10~\AA~range, which is entirely explained by the nanopores in the surface. A typical cross section of such a pore, containing a CO molecule, is shown in Figure~\ref{fig:co_pore}. The snapshot was taken from the AKMC simulation of CO on substrate $S_1$. The position of this cross section is also indicated in Figure~\ref{fig:height_surface}. The deeper well in the upper left corner on the crystalline substrate is not a special site but is an artifact of the mapping method. It corresponds to one of the hexagons in the ice Ih crystal which has a grid point almost exactly above its center, making it appear deeper than the other hexagons. All the hexagons in ice Ih are too small for CO to diffuse into. From the surface height map, the surface area of the substrates can also be calculated by performing a polygonal triangulation. Again, little variation between the amorphous substrates was found. The average area is $807\pm7$~\AA$^2$, which is $1.31\pm0.01$ times the area of the base of the simulation box. For ice~Ih, this factor is $1.13\pm0.02$. 

We want to stress that the pores in our simulated ices are of nanometer size and should not be associated with the porosity in laboratory ices~\citep{Bossa2012}. The laboratory pores are much larger and define the structure of the ice on the macroscale. Our nanopores can fit at most two CO molecules whereas the experimental pores are typically assumed large enough to fit ten to hundreds of molecules. For clarity, we will refer to the pores in experimental ices as macropores throughout this paper. Of course, nanopores are also expected to be present\new{ on the surface of laboratory ices, but also on the walls of the bulk macropores. Hence, physical effects arising from the presence of nanopores in the simulations will also be present in laboratory ices.}

\subsubsection{AKMC Simulations and Binding Sites}
\label{sec:binding}
We performed AKMC simulations at $T=50$~K on all three substrates, starting from a configuration with a single CO admolecule relaxed on the substrate\new{,} at a random position. \new{The temperature of 50~K was chosen because, when compared to lower temperatures, it reduces the number of KMC steps necessary to explore all binding sites on the surface. The TOEs are then readily adjusted for simulations at lower temperatures.} While the simulation explores the PES and the system evolves in time, not only does the admolecule diffuse, but the substrate itself also evolves. In contrast with our previous calculations on crystalline ice~\citep{Karssemeijer2012}, we observe here that the water molecules in the amorphous substrate are rather \new{mobile}. Many states are found where the CO molecule remains in roughly the same position but where the water molecules have moved enough for the states to be considered distinct. This is a manifestation of the much rougher and more complicated PES, which has many more shallow minima compared to crystalline systems. The hydrogen bond network, however, remains mostly unchanged during the simulation.

This roughness has an important consequence for the simulations. On crystalline substrates, the number of states entered by the simulation remained limited since the water molecules did not move significantly and the CO molecule could only occupy a finite number of binding sites. For the amorphous systems this is not the case; since the water molecules also move, the simulation keeps entering new, unexplored states and expensive saddle point searches are almost continuously required. We therefore had to stop the AKMC sampling manually, once confident that a sufficiently large set of states was explored.

\begin{figure*}[t!]
\includegraphics{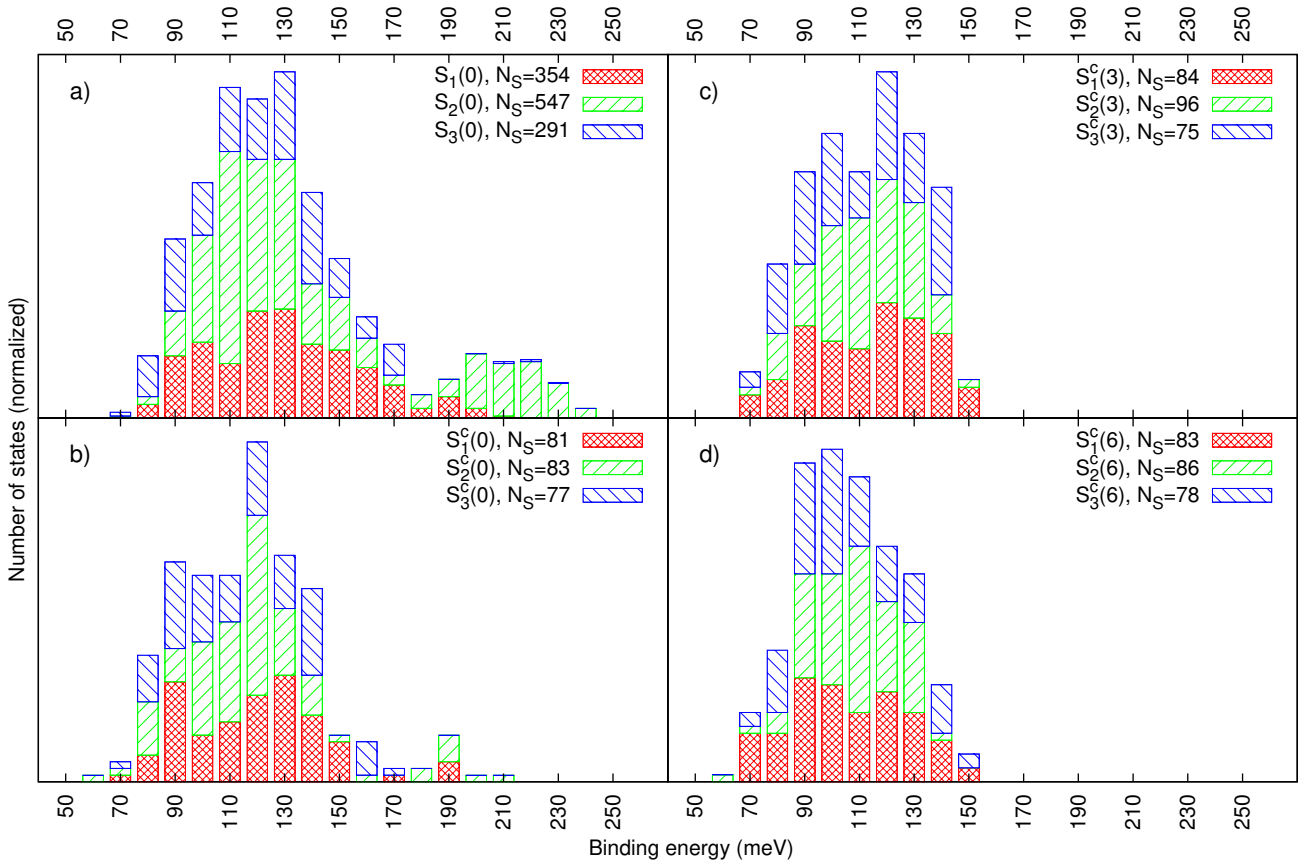}
\caption{Distribution of CO binding energies on the different amorphous ice substrates studied. Panel a) contains the free substrates with one admolecule. Panels b), c), and d) show binding energies on constrained substrates with 0, 3, and 6 additional CO molecules, respectively. The classification of the substrates is explained in Section~\ref{sec:compdetails}; $N_s$ is the number of states explored by the AKMC simulations on each substrate.\label{fig:be_hist}}
\end{figure*}

AKMC simulations were also performed on the constrained substrates ($S_1^c$, $S_2^c$, and $S_3^c$), where the number of states is limited due to the frozen substrate (now only the CO molecule is allowed to move). Here, we found about 80 states per substrate (see Figure~\ref{fig:be_hist}). To be sure that we sampled long enough on the unconstrained samples, we visually verified that the binding sites found on the constrained substrates are also present on the corresponding unconstrained sample.

For all states entered by the AKMC simulations, we calculated the binding energy, $E_{\text{B}}$, of the CO molecule to the ice surface by taking the difference between the energy of the system with the adsorbed CO and the energy of the substrate, after relaxing it without the CO. The distribution of binding energies and the number of states on both the free and the constrained substrates are shown in the left panels of Figure~\ref{fig:be_hist}. The binding energies are found to be broadly distributed between 60 and 250~meV, much broader than the distribution for crystalline ice (between 100 and 150~meV) we found in our previous study. 

\begin{figure}
\includegraphics{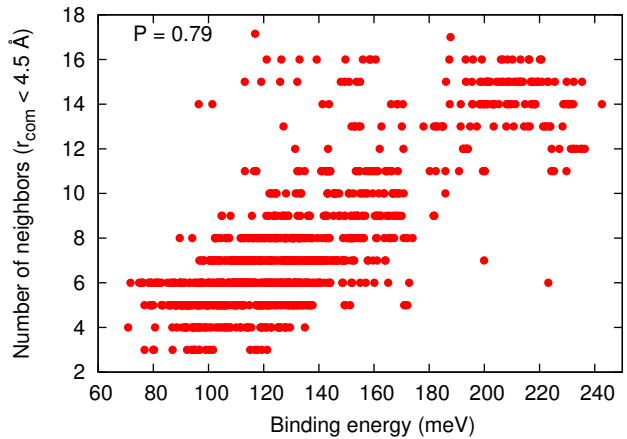}
\caption{Correlation between the binding energy and the number of neighbors for a single adsorbed CO molecule on ASW.\label{fig:scatter}}
\end{figure}

The origin of the broad distribution of binding energies is revealed by correlating, for each state, the binding energy with the number of H$_2$O molecules neighboring the CO. We defined the number of neighbors as the number of H$_2$O molecules which have their center of mass within a radius 4.5~\AA~of the center of mass of the CO molecule. The correlation with the binding energy is shown in Figure~\ref{fig:scatter} and has a P-value of 0.79. Physically, this means that the nanopores in the ASW substrate provide the strongest binding sites, which is consistent with experimental observations of trapped CO in ASW~\citep{Devlin1992,Collings2003b}. Previous theoretical investigations of the adsorption of CO on ASW report a similar correlation between the CO binding energy and its number of neighbors, though the reported maximum binding energy of 155~meV is significantly lower than our highest binding energy~\citep{Al-Halabi2004a}. A possible explanation for this, besides the different interaction potential, is that in this work, the binding energy was found from geometry optimizations starting from the final configurations of 15~ps molecular dynamics trajectories. As we shall see at the end of  Section~\ref{sec:singledynamics}, this may not leave sufficient time for the adsorbate to find its way into one of the pores, where the binding energy is highest. 

The analysis above shows that when classifying the binding sites on the surface a good first criterion is whether the site is a pore site or a surface site. The pore sites have a higher binding energy, more neighbors, and, as we shall see in the next section, critically influence the mobility of adsorbed CO. This criterion could be used as an improvement in astrochemical models, for example by including two CO populations.

\subsubsection{Single Admolecule Dynamics}
\label{sec:singledynamics}
The complicated structure of the substrate and the PES also influences the efficiency of the AKMC simulations to simulate long-timescale diffusive behavior. One aspect is that the roughness of the PES leads to many states, separated by only low barriers. It is a known problem that the KMC algorithm tends to get stuck in these states, crossing and recrossing a low barrier many times before eventually evolving over a higher, physically more interesting barrier. As we demonstrated before for ice Ih substrates, this effect can be countered by using a coarse graining algorithm~\citep{Pedersen2012a}. We employed this algorithm in all AKMC simulations presented here. 

A second aspect is that the admolecule resides much longer in the regions of the substrate with nanopores than in the regions with just surface sites, which becomes clear by analyzing the time spent by the admolecule on the different parts of the substrate. Especially for systems with an unconstrained water substrate, this gives rise to a problem. In this case, we observe in our simulations that the CO molecule is able to diffuse into one of the pores quickly, but once it is there, many different states are found by the AKMC method where the CO molecule remains more or less in the same position but where all molecules, including the surrounding H$_2$O molecules reorient themselves somewhat, in a concerted motion. One should be mindful of this effect of the nanopores when looking at the histograms of binding energies on the unconstrained substrates in Figure~\ref{fig:be_hist}. Even though the number of physical pores on each substrate is limited (typically about three per substrate), each of them contains a large number of states. As for the KMC simulations, these states inside a pore often have low barriers between them, making it difficult for the simulation to escape from a pore region in a reasonable \new{number} of KMC steps. For this reason we were unable to extract a reliable diffusion coefficient on the unconstrained substrates. Only on $S_3(0)$ were we able to make an estimate of $6.4\pm3.4\times10^{-13}$ cm$^2$s$^{-1}$ at $T=50$~K.

\begin{figure}[h]
\includegraphics{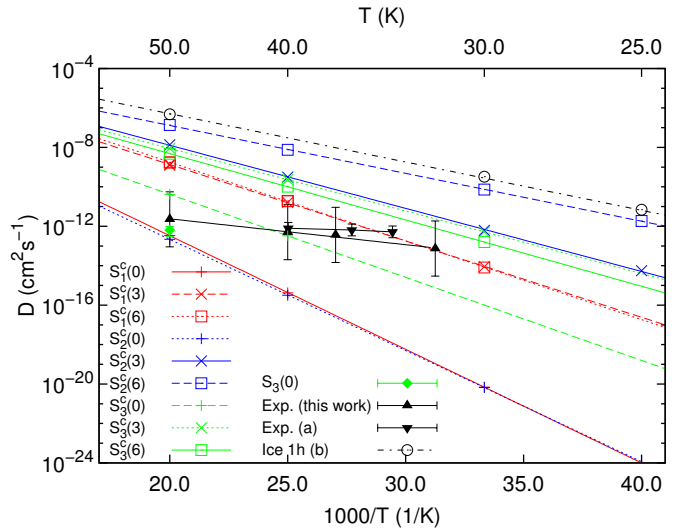}
\caption{CO diffusion coefficient on frozen substrates with varying coverage. \new{For comparison, the green triangle shows the diffusion coefficient on the unconstrained substrate, $S_3(0)$.} Experimental data from the isothermal desorption experiments is also shown, as well as experimental results from~\cite{Mispelaer2013} (a). The value for ice Ih is that of substrate sample 1 from \cite{Karssemeijer2012} (b).\label{fig:diffusion_frozen}}
\end{figure}

On the constrained substrates this effect from the nanopores is much less prohibiting since the substrate molecules cannot move to contribute to a large number of shallow pore states. This is the reason why Figure~\ref{fig:be_hist} shows fewer sites with high binding energies on the constrained than on the unconstrained substrates. The CO is able to enter the pores on the constrained samples, but the water molecules cannot reorient themselves to accommodate for the CO molecule, whereby increasing the binding energy and the number of states. We performed KMC simulations and extracted the diffusion coefficients on all three constrained substrates in the temperature range between $25$ and $50$~K. These results, \new{as well the value for the unconstrained substrate $S_3(0)$}, are shown in Figure~\ref{fig:diffusion_frozen}.

The diffusion coefficient at $T=50$~K for the constrained substrate $S_3^{c}(0)$ is more than one order of magnitude higher than the value for the unconstrained substrate. Even though the latter is not as reliable due to the bad statistics of the KMC simulations, this discrepancy is consistent with comparisons we performed between CO diffusion on constrained and unconstrained ice Ih substrates. Also, \cite{Batista2001} report a lower diffusion energy barrier for water surface diffusion on constrained ice Ih surfaces than on freely moving substrates. This is because substrate relaxations lower the potential energies of minima on the PES more than at the saddle points. \new{For our systems, the process barrier heights on the unconstrained substrates are on average  10\% higher than on their constrained counterparts. Based on this rough estimate, the diffusion coefficient at  50~K, assuming a barrier height of 100~meV, leads to a ten times higher CO diffusivity on the constrained samples.} For this reason, the diffusion coefficients on the constrained substrates given in this paper should be interpreted as upper limits for the true diffusivity of the completely free system.

Figure~\ref{fig:diffusion_frozen} clearly shows that the diffusion coefficient, $D$, of a single CO on the constrained substrates is described by an Arrhenius expression 
\begin{equation}
D(T) = D_0 \exp{\left(-\frac{E_{\text{D}}}{k_{\text{B}} T}\right)}.
\end{equation}
By fitting the equation to our data we extracted the effective diffusion activation energies, $E_{\text{D}}$, and the pre-exponential factors, $D_0$, for all three substrates. These values are listed in Table~\ref{tab:diffusion}.

For reference purposes, we also show our earlier results on the diffusivity of a single CO molecule on ice Ih. On the crystalline substrate, the diffusion coefficient is at least four orders of magnitude larger than on any of the constrained amorphous substrates. Regardless that the diffusivities on the constrained amorphous substrates should be considered as an upper limit, it is clear that the single CO mobility on an amorphous ice substrate is very low. Our KMC simulations show that, even at $T=50$~K on $S_3^{c}$ (the substrate with the highest mobility), the admolecule spent 98\% of the simulation time trapped in one of the pores. So if, in a molecular cloud, a CO molecule lands on an H$_2$O dominated ice mantle, it will be mobile for only a short time, until it reaches a strong binding pore site. Averaged over all three unconstrained substrates, the time it takes the CO to reach a pore is 7 ns at $T=50$~K. We determined this time by averaging KMC trajectories which we started from a random weakly bound surface site ($E_{\text{B}}<100$~meV) and stopped once a strong binding site was entered ($E_{\text{B}}>150$~meV). Although 7 ns is a short time in our simulations, it is almost two orders of magnitude longer than the classical trajectory calculations by \cite{Al-Halabi2004a}. This might explain why we find sites of higher binding energy in our simulations.

\new{Because the average binding energy is significantly larger than the effective activation energy for diffusion, desorption will occur on a much longer timescale than surface diffusion. For example: at $T=50$~K, a binding energy of 150~meV and a diffusion barrier of 100~meV leads to a timescale difference of five orders of magnitude between diffusion and desorption. This difference only increases for lower temperatures. Consequently, the absence of desorption as a process in our TOEs is not a problem for our mobility analysis.}

\subsubsection{Filling the Pores}
The simulations of a single adsorbed CO on amorphous ice have revealed that the porous nature of the substrate effectively immobilizes the admolecule. However, both in molecular clouds and in laboratory experiments, there are multiple CO molecules. Given the previous section, some of these molecules will fill the nanopores which naturally raises questions regarding the mobility of the remaining CO molecules, which are not trapped. 

We address this question by occupying the three, or six, strongest binding sites on each substrate with CO molecules. This corresponds to filling the nanopores with adsorbates. These configurations  were then relaxed and this entire system is constrained from movement. Then we added one more CO admolecule, free to move, which we studied by means of AKMC. These systems with three or six additional CO molecules are denoted respectively by $S_i^c(3)$ and $S_i^c(6)$ where $i=1,2,3$ labels the substrate.

For these new substrates, we follow the same procedure as before. The number of explored states and the corresponding distribution of binding energies is shown in the right panels of Figure~\ref{fig:be_hist}. The successive disappearance of the accessible high binding energy sites, as the coverage increases, is clearly observed in the distribution, whereas the lower energy side of the distribution remains relatively unchanged. The effect on the mobility is even more pronounced. As shown in Figure~\ref{fig:diffusion_frozen}, the diffusivity on the new substrates is greatly enhanced with respect to the bare ices, even though it does not exceed the value on hexagonal ice. This increased mobility results in significantly lower diffusion barriers, as can be seen in Table~\ref{tab:diffusion}. It is worth noting that the effect of going from zero to three occupied sites is much larger than that of going from three to six. This is explained by the surface height analysis in Sec.~\ref{sec:morphology}. This shows that none of the substrates has more than three `real' nanopores making the step from three to six occupied sites much less dramatic. Based on the diffusion barriers one could even argue that the effect is negligible, given the range of barriers over the three different substrates, which can of course be considered as being three distinct surface regions on a larger amorphous ice surface.

\begin{deluxetable}{cccc}
\tablecaption{Diffusion coefficients at $T=50$~K, and fitted Arrhenius parameters for diffusion on all CO-amorphous ice systems studied. Experimental data and values extracted from previous calculations on hexagonal ice are also listed.}
\tablehead{
\colhead{System}& \colhead{$D$ (50 K)} & \colhead{$D_0$} & \colhead{$E_{\text{D}}$}\\
\colhead{}& \colhead{(cm$^2$s$^{-1}$)} & \colhead{(cm$^2$s$^{-1}$)} & \colhead{(meV)}
}
\startdata
S$_1$(0)          & \nodata                   & \nodata               & \nodata \\
S$_1^c$(0)    & $3.3\times 10^{-13}$      & $1.1\times 10^{-1}$   & 114 \\
S$_1^c$(3)    & $2.2 \times 10^{-9}$      & $7.5\times 10^{-2}$   & 77 \\
S$_1^c$(6)    & $1.8 \times 10^{-9}$      & $4.1\times 10^{-1}$   & 79 \\
S$_2$(0)          &                  \nodata  &             \nodata   & \nodata\\
S$_2^c$(0)    & $2.2 \times 10^{-13}$     & $4.1\times 10^{-2}$   & 112 \\
S$_2^c$(3)    & $1.4 \times 10^{-8}$      & $3.2\times 10^{-2}$   & 63 \\
S$_2^c$(6)    & $1.4 \times 10^{-7}$      & $9.4\times 10^{-3}$   & 48 \\
S$_3$(0)          & $6.4 \times 10^{-13}$    &   \nodata             & \nodata\\
S$_3^c$(0)    & $4.2 \times 10^{-11}$     & $1.1\times 10^{-2}$   & 84 \\
S$_3^c$(3)    & $9.3 \times 10^{-9}$      & $1.9\times 10^{-2}$   & 63 \\
S$_3^c$(6)    & $5.4 \times 10^{-9}$      & $2.6\times 10^{-2}$   & 67 \\
\\
Exp. (this work)          & $2.4\times 10^{-12}$    &  $9.2\times 10^{-10} $ & $26\pm15$ \\
Exp.\tablenotemark{a} & \nodata&\nodata  & $10\pm 16$ \\
Ice Ih\tablenotemark{b}                  & $4.8 \times 10^{-7}$      & $4.1\times 10^{-2}$   & 49
\enddata
\tablenotetext{a}{Experimental data from \cite{Mispelaer2013}}
\tablenotetext{b}{Theoretical value from \cite{Karssemeijer2012}}
\label{tab:diffusion}
\end{deluxetable}

\subsubsection{Higher Coverages}
\label{sec:highercoverages}
In dense cloud conditions, CO typically freezes out on grains which are already covered with an H$_2$O dominated ice~\citep{Oberg2011}. It is therefore interesting to consider layered ices which already have considerable CO buildup. Also, from an experimental point of view, simulating the dynamics of just a few admolecules is hardly representative. In TPD experiments, the surface coverage is typically much higher (ranging from 0.1 to many monolayers) and the measured molecules are those which are not trapped in the pores.

\begin{figure}[h]
\includegraphics{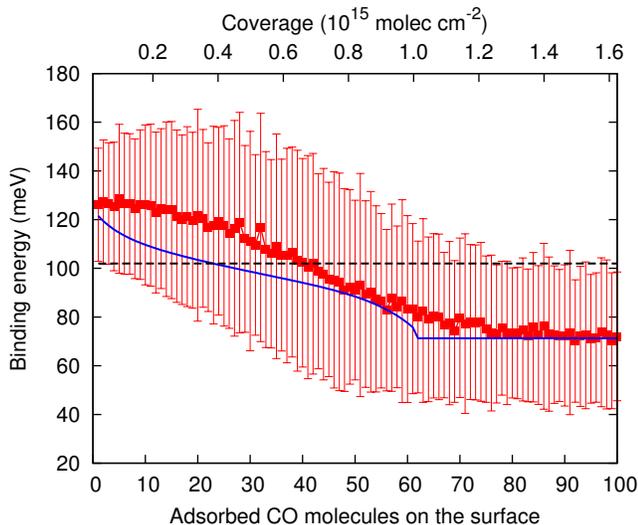}
\caption{Binding energy of CO on ASW as a function of surface coverage. The solid (blue) line shows the coverage dependent binding energy fitted to sub-monolayer TPD results by~\cite{Noble2012a}, the dashed (black) shows the value~\cite{Collings2003} extracted from TPD experiments. One monolayer is defined as $10^{15}$ molec~cm$^{-2}$.\label{fig:coverage}}
\end{figure}

For these reasons we have studied the adsorption energy of CO on amorphous ice substrates when they are already partially covered with CO. Starting from the bare water substrate $S_1$, we generate a grid at a distance of 3~\AA\ above the surface with a lattice spacing of roughly 2~\AA. A CO molecule is then placed at a randomly chosen grid point with a random orientation, the geometry is relaxed, and the binding energy of the CO is registered. Next, a new grid is generated and a second CO molecule is deposited in the same way. The energy difference between the relaxed configurations with one and two CO molecules is registered as the binding energy of the second CO molecule. In this way, 100~CO molecules were deposited and the whole procedure was repeated 450 times. The average binding energy and its standard deviation are shown in Figure~\ref{fig:coverage}. The surface coverage is also shown in monolayers, defined as $10^{15}$ molec~cm$^{-2}$, which is the commonly-used definition. Visual inspection reveals however, that one  monolayer for this system corresponds more accurately to $0.65 \times 10^{15}$ molec~cm$^{-2}$, or 40~CO molecules on the surface. Because we have about three nanopores on each substrate, about 10\% of CO can be trapped in pores at monolayer coverage.

Because our AKMC simulations show no significant diffusion of CO at low temperature and coverage, this method may well represent the experimental situation where a CO molecule is deposited from the gas phase at a random position and immediately sticks in the position where it landed on the ice. This is of course under the assumption that the CO molecules carry no significant kinetic or internal energy which would allow them to diffuse over the surface until they thermalize. In principle the method could be extended by including a certain period of equilibration, either by AKMC or molecular dynamics, after deposition.

The calculated binding energies show a large variation. This is mainly because the CO molecules are deposited at random positions, and they therefore probe a set of binding sites which, as we know from Section~\ref{sec:binding}, has a broad distribution of binding energies. Furthermore, the minimization after deposition of the $n$'th molecule may also trigger restructuring of the previous $n-1$ deposited molecules and, in theory, also of the substrate molecules, since they are also allowed to move. This would add to the binding energy of the $n$'th molecule and could lead to a systematic overestimation of the binding energy but we could not verify this. Despite the wide distribution, the mean binding energy follows a smooth trend and shows the decrease in binding energy with the surface coverage. This is because the probability of a new molecule finding a strong binding site on the substrate decreases with increasing coverage as the number of the relatively weak CO\sbond CO interactions grows while the stronger H$_2$O\sbond CO interactions decrease in number. 

It is seen that the average binding energy drops from about 125 meV at zero coverage to about 75 meV when there is already one monolayer of CO present. This is in good agreement with TPD experiments by \cite{Collings2003} who report an activation energy of 100~meV for the desorption of CO, at monolayer coverage, directly from a non-porous ASW surface.  From more recent, sub-monolayer, desorption experiments by \cite{Noble2012a} the coverage dependence of the binding energy could be extracted from the TPD data. This dependence is also well reproduced by our calculations. Data from both experiments are also shown in Figure~\ref{fig:coverage}.

\section{EXPERIMENTS}
\label{sec:experiments}
Experimentally, we determine the  diffusion coefficient of CO in an amorphous water ice environment from the rate with which CO desorbs, after it has traveled through a layer of ASW. This is achieved by depositing a slab of mixed H$_2$O:CO ice onto a substrate, followed by a layer of pure H$_2$O. This system is then kept at constant temperature while the amount of CO in the ice is monitored using infrared (IR) spectrometry. The desorption rate of CO from the ice depends on its diffusivity in the H$_2$O overlayer. The experimental procedure closely resembles the method used by~\cite{Mispelaer2013}. In our experiments however, we study a broader temperature window (32 to 50~K) and use a H$_2$O:CO mixture covered by an ASW layer, instead of a pure CO ice to avoid the diffusion of CO upon deposition of the water overlayer.

\subsection{Experimental Details}
The experiments were performed with the Caltech astrochemical ice spectroscopy setup which is described in detail by~\cite{Allodi2013}. It consists of a high vacuum chamber (base pressure $< 10^{-8}$ Torr) containing a silicon substrate which can be cooled down to $8$~K using a closed-cycle helium cryostat. Gas mixtures can be prepared in a separate metal deposition line to be deposited onto the substrate. The IR spectra of the samples are recorded by means of a Fourier Transform-IR spectrometer in transmission mode \new{at a resolution of 0.5~cm$^{-1}$.}

\subsubsection{Experimental Procedure}
The isothermal desorption experiments were performed at $32$, $37$, $40$, and $50$~K starting from an ice which was prepared as follows. First, a deposition of a mixture of H$_2$O:CO is carried out at $T=10$~K through a $1/8''$ diameter stainless steel pipe that faces the substrate. The end of the pipe is positioned about $1''$ away from the substrate and is capped with a metal mesh with a 38 micron hole size to ensure a uniform ice deposition. To remove the most loosely bound CO, this ice is then annealed twice to $32$~K at a rate of 5~K~min$^{-1}$, cooling back down, \new{with no additional waiting time}, to $8$~K in between. This creates a CO-rich ASW film with an H$_2$O:CO mixing ratio of around 2:1. Finally, an additional H$_2$O layer is deposited at 8~K to form a porous overlayer. The ice is then heated to the desired temperature at $10$~K~min$^{-1}$, and the IR spectra are recorded while the CO diffuses through the ice and desorbs from the surface. The moment when the ice reaches the desired temperature is taken as $t=0$ in the analysis of the results.

The column densities of H$_2$O and CO in the ice are monitored by integrating the characteristic bands of the molecules and dividing by the band strengths of the peaks. For water we use the 1660~cm$^{-1}$ bending mode and for CO the 2139~cm$^{-1}$ stretching mode. The corresponding band strengths are $1.2\times 10^{-17}$ and $1.1\times 10^{-17}$ cm molec$^{-1}$, respectively~\citep{Gerakines1995}. The area of the water band was determined by numerical integration while the area of the CO band was calculated by fitting two Gaussians to the spectrum to distinguish between CO in water-rich (polar) and water-poor (apolar) environments. The polar component has a distinct peak around 2152~cm$^{-1}$ while the larger, non-polar, component peaks around 2139~cm$^{-1}$ \citep{Sandford1988,Bouwman2007}.

We estimate the thickness of the ice from the column densities using densities of $0.94$~g~cm$^{-3}$ for H$_2$O~\citep{Jenniskens1994} and $0.81$~g~cm$^{-3}$ for CO~\citep{Loeffler2005}. Before the start of the isothermal experiments, the ice consists of a $1.7\pm0.4$~$\mu$m thick mixed layer with an $0.38\pm0.03$~$\mu$m layer of pure H$_2$O on top. This ratio was similar for every temperature and is in good agreement with the deposited amounts of gas (3~Torr of mixture followed by 1~Torr of H$_2$O) as measured in the dosing line by a mass independent active capacitance transmitter. We are therefore confident that the sample we start from is the same for all experiments. \new{Furthermore, by performing the annealing cycles before depositing the H$_2$O overlayer, the most weakly bound CO molecules desorb. This amounts to about 4\% of the total CO. The annealing procedure limits diffusion of CO into the H$_2$O cap during its deposition and makes the ices at the start of each isothermal experiment more similar.}

After the isothermal experiments, a TPD experiment is performed with a heating of 1~K~min$^{-1}$, until the ice has fully desorbed. IR spectra are taken every minute during the heating.

\subsubsection{Determination of the Diffusion Coefficient}
The concentration profile of CO, $n(z,t)$, in the ice is described using a solution to Fick's second law of diffusion in one dimension:
\begin{equation}
\frac{\partial n(z,t)}{\partial t} = D(T)\frac{\partial^2 n(z,t)}{\partial z^2}\label{eq:Fick2}.
\end{equation} 

This approach was shown to give good results by \cite{Mispelaer2013}. The solutions of this equation depend  on the initial conditions. For our purpose we impose that $n(h,t)=0$ to reflect immediate desorption of molecules which reach the surface (at $z=h$) and $\partial n(0,t)/\partial z = 0$ because no CO can escape from the bottom of the film. Furthermore, the concentration profile at $t=0$ is chosen to be either $n_{\text{s}}$ or $n_{\text{h}}$:

\begin{align}
    n_{\text{h}}(z,0) &= {\phantom{\Big\{}} \begin{array}{ccc} n_0, & \text{if } & 0 < z < h, \\ \end{array}\\
    n_{\text{s}}(z,0)    & = 
    \Big\{ \begin{array}{ccc}
        n_0, & \text{if } & 0 < z \le d, \\
        0,   & \text{if } & d < z < h.
    \end{array}
\end{align}

The first function, $n_{\text{h}}$, is used when CO is homogeneously distributed in the ice by the time it reaches the desired temperature. The second expression, $n_{\text{s}}$, describes the situation where CO is initially confined to a slab of height $d$ at the bottom of the  ice film. For these constraints the solutions to Fick's second law read
\begin{align}
n_{\text{h}}(z,t) &= \sum_{i=0}^{\infty} \frac{2n_0 (-1)^{i}}{\mu_i h}\cos{\left( \mu_i z\right)}\exp{\left(-\mu_i^2 D t\right)},\\
n_{\text{s}}(z,t) &=  \sum_{i=0}^{\infty} \frac{2n_0}{\mu_i h}\sin{\left(\mu_i d\right)}\cos{\left( \mu_i z\right)}\exp{\left(-\mu_i^2 D t\right)},
\end{align}
for $n_{\text{h}}$ and $n_{\text{s}}$ respectively. Here $\mu_i=(2i+1)\pi/2 h$ and $D$ is the diffusion coefficient. From these expressions\new{,} the column density of CO molecules in the ice is readily found by integrating over $z$ from $0$ to $h$. These can be converted to band areas, $A_{\text{h}}$ and $A_{\text{s}}$, which can then be fitted to the experimental data. The final expressions are
\begin{align}
A_{\text{h}}(t) &= s+\sum_{i=0}^{\infty} \frac{2(A_0-s)}{\mu_i^2 h^2}\exp{\left(-\mu_i^2 D t\right)}\label{eq:A_cont},\\
A_{\text{s}}(t) &= s+\sum_{i=0}^{\infty} \frac{2(A_0-s)(-1)^i}{\mu_i^2 hd}\sin{\left(\mu_i d \right)}  \exp{\left(-\mu_i^2 D t\right)}\label{eq:A_slab},
\end{align}
where $A_0$ is the initial band area and $s$ is an offset which we have to include to reproduce the experimental data. Physically this corresponds to CO which is completely trapped in the water ice and cannot diffuse out.

\new{More complicated models, with separate diffusion coefficients for CO in the upper and lower layers, were also tried. With these models, better fits to the data could be obtained, but we did not find two clearly distinguishable diffusion coefficients. We therefore attribute the better fit to the increased number of parameters in the model and decided to stick with the simpler models of Eqs~\eqref{eq:A_cont} and~\eqref{eq:A_slab}.}

\subsection{Experimental Results and Analysis}
\begin{figure}[h]
\includegraphics{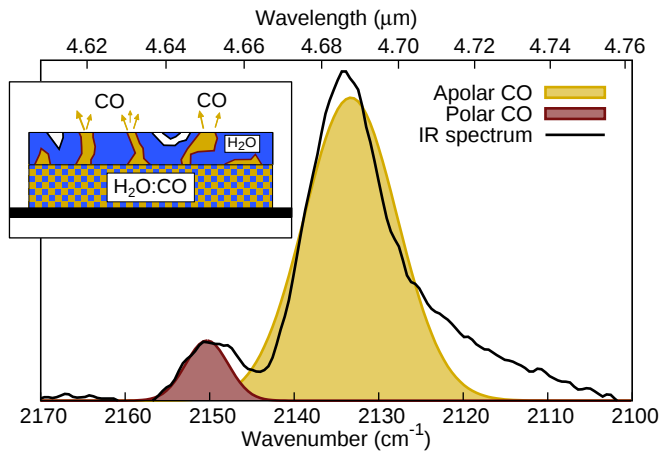}
\caption{Infrared spectrum of the CO peak taken at $t=1$~hr. in the isothermal desorption experiment at $T=37$~K. The inset schematically shows the structure of the ice\label{fig:spectrum}.}
\end{figure}

\begin{figure}[h]
\includegraphics{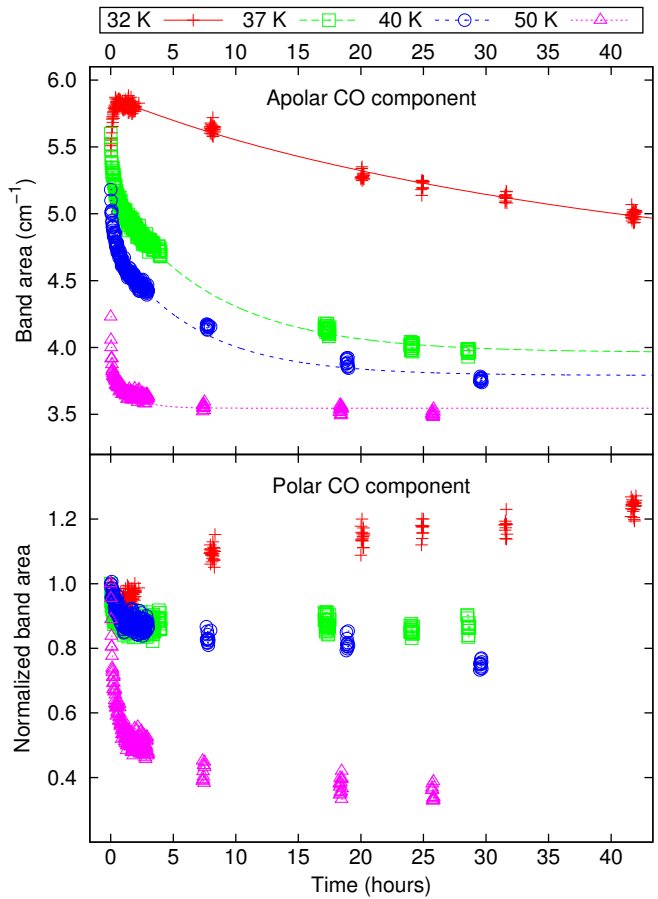}
\caption{Top panel: band area of the apolar CO component (points) and model fits (lines) as a function of time as measured during the isothermal desorption experiments. Bottom panel: normalized area of the polar CO band\label{fig:FitD_polarCO}.}
\end{figure}

In the isothermal desorption experiments, the IR spectra are recorded once the deposited ice reaches the desired temperature. \new{The spectrum of the CO peak, recorded after 60~minutes at $T=37$~K, is shown in Figure~\ref{fig:spectrum}. At this time, CO is diffusing through the porous ASW overlayer and desorbing from the surface. This is schematically shown in the inset of the figure. The two fitted Gaussians, which were used to estimate the CO band area, are also shown. The time evolution of the band areas corresponding to the polar and apolar peaks, are shown} in Figure~\ref{fig:FitD_polarCO}. The polar component is always just a small fraction ($6\pm 3 \%$) of the total CO stretch band area and to show its behavior we have normalized it to its $t=0$ value. Due to the small contribution from the polar band, the total CO band area is almost identical to the apolar component and it decreases in time, due to desorption from the ice. The decay times are seen to decrease with increasing temperature which we attribute to faster diffusion of CO through the ASW overlayer. \new{At $T=32$~K, an increase is seen in the band area of the apolar peak during the first $\sim$30~minutes of the experiment. We believe that this is due to changes in band strength arising from structural changes in the ice. These include the dilution due to diffusion of CO into the upper layer and possibly the local crystallization of CO. At 32~K we think this process is slow enough, in combination with the low CO desorption rate, to be observed in the IR spectra. At higher temperatures, the changes will also occur, but these then proceed too rapidly to be observed.}

\new{The behavior of the polar component, corresponding to CO interactions with dangling OH bonds, differs from that of the apolar component. As seen from the lower panel of Figure~\ref{fig:FitD_polarCO}, the polar component  only decreases significantly at $50$~K, the highest temperature. At the lower temperatures, the polar component remains almost constant, while the total amount of CO decreases. At $T=32$~K, the polar component even increases with time. This suggests that the `polar CO population' is less mobile than the apolar CO and thus corresponds to CO occupying strong binding sites within the ASW. The increase at $32$~K is then naturally explained because, as diffusion into the upper layer progresses, more of these energetically favorable sites become available for the CO, leading to an increase of the polar band. From the spectra taken during the TPD, following the $32$~K isothermal experiment, it seems that the polar component decreases most rapidly around $40$~K, consistent with the data from Figure~\ref{fig:FitD_polarCO}. Unfortunately, the data is too noisy to draw definitive conclusions and the analysis of the IR spectra during the TPD remains speculative.}

From the analysis above, we conclude that the apolar CO band is the best measure of the mobile molecules and we thus use the data from this component to fit our Fickian model. The band area contribution from the polar peak is omitted from the fit because it is a measure of molecules which are generally bound too strongly to diffuse. The rapid initial decrease in band area at temperatures of $37$~K and higher suggest that CO has already diffused into the upper H$_2$O layer at $t=0$. This means that CO in homogeneously distributed in the ice at the start of the isothermal measurements, so we use Eq~\eqref{eq:A_cont} to fit the data in these cases. At $T=32$~K there is an incubation period of about one hour before the CO band area starts decreasing. This behavior indicates that the CO is still confined to the lower layer at $t=0$ and we therefore use Eq~\eqref{eq:A_slab} as a model for this temperature. \new{Because the model cannot describe the increasing band area during the first 30~minutes at this temperature, these data were excluded from the fit.}

To fit the data, the diffusion coefficient $D$, the offset $s$, and the initial band area $A_0$ were used as fitting parameters. The fit of the initial band area was needed because there is too much noise in the data to keep it fixed at the measured value at $t=0$. The thickness of the CO containing slab $d$ and of the total ice film $h$ were fixed parameters determined from the spectra taken right after the second annealing of the ice and after deposition of the H$_2$O overlayer, respectively. All parameters, including the diffusion coefficients, are listed in Table~\ref{tab:experimentalresults} and the Arrhenius behavior of the diffusion is shown in Figure~\ref{fig:diffusion_frozen}. From the latter we extract a diffusion energy barrier of $26\pm 15$~meV.

From Figure~\ref{fig:FitD_polarCO}, we see that the model is able to describe the experimental data with reasonable accuracy. Especially the incubation time at $T=32$~K is well described by the solution with the CO initially confined to the lower part of the ice. Despite the relatively good fit there are several effects which lead to large uncertainties in the extracted diffusion coefficient. First, some uncertainty arises from the determination of the thickness of the ice film. Our estimate depends on the densities of CO and H$_2$O which will change due to heating and, especially for water, depend strongly on the deposition method and temperature~\citep{Stevenson1999,Dohnalek2003}. The thickness estimate also depends on the band strengths, which are also influenced by temperature and by the mixing ratio~\citep{Bouwman2007}. We believe that the thickness is the most uncertain parameter in the model and the error bars in Figure~\ref{fig:diffusion_frozen} are based on a 50\% uncertainty of the ice thickness. This however, leads to a systematic error which affects the pre-exponential factor, $D_0$, but not energy barrier $E_{\text{D}}$. A related aspect is the structural change in the water ice during the experiment which leads to compaction of the film due collapse of macropores~\citep{Bossa2012} and subsequently to trapping or release of CO~\citep{Bar-nun1985,Collings2003b}. In a pure ice, this transition is observed between $38$ and $68$~K~\citep{Jenniskens1994}. Another source of error is the manner in which the band areas are extracted from the IR spectra. \new{As can be seen from Figure~\ref{fig:spectrum}, the fit is not perfect. We attribute this mainly due to the large amount of H$_2$O in the chamber, which affects the baseline of the spectrum in the CO stretch region.} Differences between numerical integration of the bands versus the fitting of Gaussians and various methods of baseline subtraction influence the extracted diffusion energy barrier significantly. Given these considerations, we estimate the uncertainty in the diffusion barrier to be $15$~meV.

The experimental results from \cite{Mispelaer2013} are also shown in Figure~\ref{fig:diffusion_frozen}. Even though the authors also mention several sources of errors in the data, it is reassuring to see that there is good agreement between the two experiments where the temperatures overlap. We find a somewhat steeper slope on the diffusion coefficient which is reflected in a higher energy barrier of $(26\pm15)$~meV against $(10 \pm 15)$~meV from \cite{Mispelaer2013}. The key difference between the two experiments is that we start by depositing a mixture of H$_2$O and CO instead of a pure layer of CO. This procedure binds the CO molecules more strongly in the ice film and allowed us to perform isothermal experiments up to $50$~K. Additionally, the mixture of H$_2$O:CO provides a better thermal conduction with the H$_2$O overlayer which decreases the temperature gradient in the ice. The larger temperature range studied facilitates the extraction of the temperature dependence of the diffusion coefficient and the calculation of the energy barrier for diffusion.

\begin{deluxetable}{cccccc}
\tablecaption{Parameters used to model the isothermal desorption experiments.}
\tablehead{
\colhead{$T$}& \colhead{$D$ (Fitted)} & \colhead{$A_0$ (Fitted)} & \colhead{$s$ (Fitted)} &\colhead{$h$} & \colhead{$d$}   \\
\colhead{(K)}& \colhead{(cm$^2$s$^{-1}$)}  & \colhead{(cm$^{-1})$} & \colhead{(cm$^{-1}$)}& \colhead{($\mu$m)} & \colhead{($\mu$m)}
} 
\startdata
32 & $7.4 \times 10^{-14}$ & 5.8 & 4.4 & 1.72 & 1.31 \\
37 & $3.7 \times 10^{-13}$ & 5.4 & 4.0 & 1.63 & 1.28 \\
40 & $5.1 \times 10^{-13}$ & 5.0 & 3.8 & 1.73 & 1.32 \\
50 & $2.2 \times 10^{-12}$ & 3.9 & 3.5 & 1.73 & 1.37   
\enddata
\label{tab:experimentalresults}
\end{deluxetable}

\section{COMBINING SIMULATIONS AND EXPERIMENTS}
\label{sec:comparison}
The computational and experimental predictions of the diffusivity of CO in an ASW environment can be compared in both Table~\ref{tab:diffusion} and Figure~\ref{fig:diffusion_frozen}. Although the simulations describe a pure surface processes and the experiments measure bulk CO diffusion through a ASW layer, the two are comparable given the porous nature of the vapor deposited water ice, if we assume that the size of the cracks and macropores in the ASW are sufficiently large to interpret the CO diffusion as an effective surface process, along the pore walls. This assumption is reasonable because we start from a macroporous ice and the diffusion rate of CO at these low temperatures is much faster than the reorganization rate in ASW, with which the porous structure can collapse~\citep{Mispelaer2013}. \new{The diffusion of CO along the macropore walls is then comparable to the diffusion in the AKMC simulations because the walls of the macropores will also contain nanopores, similar to those on the simulated ices.} On a qualitative level, we see that there is good agreement between the simulations and experiments on two key points. First and foremost, the diffusion coefficients are in good agreement on an order of magnitude level. Secondly, both simulations and experiment show that a fraction of CO is immobilized by the water due to trapping in nanopores. We will discuss these points in more detail below.

The diffusion coefficients can be compared between theory and experiment but one should keep in mind that in the experiments the coverage of CO molecules is significantly higher so the mobility is more strongly influenced by the weak CO\sbond CO interactions than in the simulations. The simulations where the nanopores are filled with CO are thus the most representative of the experimental situation. This is also reflected by the flattening of the slope in Figure~\ref{fig:diffusion_frozen} when the CO coverage is increased. The diffusion barriers extracted from the simulations with 3 or 6 of the pore sites filled vary between the substrates but are all within $66\pm20$ meV. This is higher than the experimental value of $26$~meV. The stronger contribution from the CO\sbond CO interactions is one possible explanation for this discrepancy. Another effect which could play a role is, as mentioned before, the simultaneous compaction of the ASW film and the CO diffusion through it. At higher temperatures, the ice becomes more compact due to the closing of cracks and the collapse of macropores. This results in a relatively lower mobility at higher temperatures and thus a flatter slope and lower diffusion barrier. The agreement between the pre-exponential factors, $D_0$, is not as good. From the simulations, this attempt frequency derives mainly from the vibrational excitations of the system and the values we find are close to the value of $10^{-3}$~s$^{-1}$ which is often used~\citep{Kellogg1994}. From the analysis of the experimental results we have seen that the pre-exponential varies largely, though systematically, with the ice film thickness and conclusive values cannot be obtained. As input for astrochemical models, the pre-exponential factor \new{has less significance} than the energy barrier for diffusion. 

The key importance of the morphology of the water substrate is seen in both the experiments and the simulations. The simulations show clearly that pores in the ASW substrate, even if only sub-nanometer in size, have a critical influence on the mobility of adsorbed CO. For the CO to be mobile on the surface, it is a prerequisite to have either no pore sites or to have them all filled. In the experiments, we see the polar CO band, corresponding to CO interacting with OH dangling bonds,\new{ remaining constant,} or even increasing in strength\new{,} while the total amount of CO is decreasing. This indicates that this band corresponds to those CO sites with the highest binding energy. This leads us to the conclusion that the CO signal from the polar band in the IR spectra corresponds to CO occupying sites similar to the, nano-sized, pore sites in the simulations.

A final remark regarding the comparison between simulations and experiments concerns the effect of the potential energy functions used in the simulations. The somewhat higher binding energies of CO on ASW compared to the experimental value (see Figure~\ref{fig:coverage}) could point to an overbinding in our H$_2$O\sbond CO potential. Although the diffusion coefficients are not derived from the absolute value of the potential but from the energy difference between the minima and saddle points on the PES, it could be that the minima are affected more strongly than the saddle points leading to a higher diffusion energy barrier.

\section{ASTROPHYSICAL SIGNIFICANCE}
\label{sec:astrophysical}
The AKMC simulations, as well as the isothermal desorption experiments have demonstrated that nanopores in ASW play an important role in the kinetics of adsorbed CO. These pores, where an adsorbed CO molecule can interact with a large number of water molecules, are found to have very high binding energies, leading to trapping of CO at low coverages. In the Monte Carlo simulations, we observed that a single CO admolecule was trapped in one of the nanopores for 98\% of the simulation time, even at $T=50$~K. This effectively immobilized the admolecule. Diffusion became more rapid once the pore sites were effectively removed from the ice by filling them with CO molecules. In the experiments, the strong binding energy of the nanopore sites was deduced from the increased intensity of the 2152~cm$^{-1}$, polar CO, band, associated with CO interactions with dangling OH bonds in the ice.

The large influence from the morphology of the substrate on the CO mobility is also seen when comparing to our previous results on crystalline ice~\citep{Karssemeijer2012}. Simulations at the lowest coverage, with just a single CO admolecule show that the diffusion coefficient on crystalline substrates is about four orders of magnitude larger than on amorphous surfaces. The energy barrier for diffusion, $E_{\textrm{D}}$, on crystalline ice (49~meV) is 50~\% lower than in the amorphous case ($\sim100~$meV). The diffusion prefactor is however rather unaffected by the morphology of the substrate.

Given the amorphous character of interstellar dust grain mantles, the presence of small pores will be a key factor determining formation rates on grain surfaces. In the early stages of dense cloud formation, before catastrophic CO freezeout, nanopores are likely to trap carbon monoxide molecules for very long times. This will affect the surface chemistry because the pores can act as reactive sites in this case. Although the overall mobility of reactants will be low, they will tend to get trapped in the same places, giving more time for reaction to occur, which is especially favorable if there is a reaction barrier to overcome. This was found by \cite{Fuchs2009} for hydrogenation reactions in CO ices. 

\new{The diffusion barrier of hydrogen atoms on ASW is much lower than that of the CO molecules~\citep{Perets2005,Matar2008,Hama2012}. Recent experiments by~\cite{Hama2012} have shown that the majority of H atom binding sites on ASW are shallow, with diffusion barriers $\le22$~meV. A small fraction of the sites was found to have higher diffusion barriers ($\ge 30$~meV), which might correspond to nanopore sites. Based on the results of~\cite{Minissale2013}, also the diffusion barrier of atomic oxygen on ASW is lower than that of CO. This rapid diffusion of atomic hydrogen and oxygen will lead to an efficient conversion of CO, trapped in pore sites at low coverages, to CH$_3$OH, CO$_2$ and HCOOH (with OH as a potential intermediate). If this conversion is efficient enough, there will be no CO left in the pore sites of the H$_2$O dominated ices. This is consistent with the non-detection of the $2152$~cm$^{-1}$ in astronomical spectra~\citep{Pontoppidan2005} and the suggestion that CO is mixed with CH$_3$OH in dust grain mantles to account for the red component of the CO band~\citep{Cuppen2011B}}.

Once more CO freezes out on the grain, the nano-sized pores get filled and the remaining CO can diffuse much faster. At the same time however, the remaining CO molecules will also desorb more easily, because they are more weakly bound. We computationally studied this decrease in binding energy as an increasing amount of CO molecules was adsorbed on an ASW substrate and found good agreement with experimentally determined trends. This good agreement also adds to the reliability of the simulations at low coverages, where experimental data are still scarce. 

To provide astrochemical modelers with necessary input parameters, we determined the energy barrier for diffusion from both the simulations and the isothermal desorption experiments. The simulations were found to be in reasonable agreement with the experiments as well as with similar experiments carrier out by~\cite{Mispelaer2013}. Even though there are many uncertainties, these are to the best of our knowledge the only available data on CO diffusion in ASW. The analysis has shown that the CO mobility and its binding energy are highly dependent on both the position on the ice surface as well as on the CO coverage. In this respect, amorphous surfaces are essentially different from crystalline substrates, which show much less inhomogeneity. Because diffusion is an important parameter, modelers should try to include as much of these local variation as possible and avoid taking diffusion barriers as fixed ratios of the binding energy. 

There are several approaches to include effects from inhomogeneity in the ice mantle in astrochemical models. In lattice gas KMC, this can be done by making the binding energy and diffusion barrier height site-dependent to account for pore sites. This site-dependent approach was used by \cite{Cuppen2005} to account for surface roughness in simulations of H$_2$ formation on dust grain analogs. 
In rate equation and master equation methods, one could include two populations of CO molecules, similar to the approach taken by~\cite{Cuppen2011a} for H$_2$. One population would represent the immobile CO molecules in the nanopores and the other the more mobile molecules on the surface.  Based on the results presented in this paper we suggest surface binding energies of 130 and 80~meV for the strong and weakly bound populations, respectively, and diffusion energy barriers of 80 and 30~meV. Another possibility to account for inhomogeneity is to include a direct dependence on surface coverage of the diffusion barriers and binding energies. In lattice KMC models this would be in the spirit of the work by \cite{Fuchs2009}, where the sticking probability of impinging H atoms was given a dependence on the H$_2$ surface coverage.


\section{CONCLUSIONS}
\label{sec:conclusions}
We have studied the dynamics of CO in amorphous water ice environments at low temperatures by means of kinetic Monte Carlo simulations and isothermal desorption experiments. The main conclusions of this analysis are the following:
\begin{enumerate}
\item The CO mobility is highly dependent on the morphology of the ice. At the lowest coverage, the presence of nanometer-size pores in ASW  increases the energy barrier for diffusion to around 100~meV; twice the value of 50~meV for crystalline ice, which does not contain pores. 

\item The surface coverage of CO on ASW critically influences the CO binding energy, as well as its mobility. When CO coverage is increased from zero to one monolayer, the binding energy decreases from 125 to 85~meV. Simulations show that the diffusion energy barriers are lowered from around 100 to 65~meV when surface pores are filled with CO.

\item Pores of sub-nanometer size in ASW form the most favorable sites for CO. In these sites, the strong binding energy leads to trapping of part of the CO population. We estimate that for an ASW substrate, covered with one monolayer of CO, about 10\% of the CO will be trapped.

\item Large scale astrochemical models can be improved by taking the effects from the molecular level, such as inhomogeneity and surface coverage, into account. We suggest including two populations of CO in rate equation models. The first population resides in the pores and strongly binds to the ASW while the second population is more weakly bound and more mobile. These two populations have binding energies of 130 and 80~meV, and diffusion barriers of 80 and 30~meV, respectively.

\end{enumerate}

\acknowledgements 
We gratefully acknowledge A. Pedersen for valuable help with the EON code and P. Theule for stimulating discussions. This work has been funded by the European Research Council (ERC-2010-StG, Grant Agreement no. 259510-KISMOL), the VIDI research programme 700.10.427, financed by The Netherlands Organization for Scientific Research (NWO), the National Science Foundation CRIF:ID and CSDM programs and the NASA Exobiology and Laboratory Astrophysics programs. M.A.A was supported by the Department of Defense (DoD) Air Force Office of Scientific Research, National Defense Science and Engineering Graduate (NDSEG) Fellowship, 32 CFR 168a. S.I. acknowledges support from a Niels Stensen Fellowship and a Marie Curie Fellowship (FP7-PEOPLE-2011-IOF-300957). L.J. Karssemeijer thanks COST Action Number CM0805 (The Chemical Cosmos: Understanding Chemistry in Astronomical Environments) for funding of several stimulating conferences and collaboration visits.

\appendix

\section{INTERACTION POTENTIALS}
\label{app:cocopot}
This appendix describes the three interaction potentials used in this work. All molecules are fully flexible and their internal motions are described by intramolecular potentials. Intermolecular interactions contain an electrostatic contribution from a set of point charges on each molecule and several functions describing the van der Waals contributions.
For the H$_2$O\sbond H$_2$O interactions, the TIP4P/2005f potential was used~\citep{Gonzalez2011}. This model was developed as a flexible version of the successful TIP4P/2005 potential which gives a good description of the condensed phases of water~\citep{Abascal2005,Vega2011}. 
The H$_2$O\sbond CO potential, as well as the CO intramolecular potential are described in \cite{Karssemeijer2012}. The intermolecular CO\sbond CO potential is described below. Because it has not been published before, we have included the details of the fitting procedure. For the sake of completeness, we also give the intramolecular part of the potential again. 

\subsection{CO-CO Potential}

\begin{deluxetable}{lccc}
\tablewidth{0pt}
\tablecaption{Potential parameters for the CO\sbond CO Buckingham potential.}
\tablehead{
\colhead{Interaction} & \colhead{$A_{ij}$} & \colhead{$B_{ij}$} & \colhead{$C_{ij}$} \\
\colhead{} & \colhead{(eV)} & \colhead{(\AA$^{-1}$)} & \colhead{(eV\AA$^{6}$)} 
}
\startdata
C-C & 361.4 & 2.835 & 33.45 \\ 
C-O & 1517 & 3.543 & 15.19 \\
O-O & 6370 & 4.252 & 10.55
\enddata
\label{tab:table1}
\end{deluxetable}

Following up on the work by \cite{Vissers2003}, the potential energy surface of the CO dimer was calculated from a set of interaction energies on a grid of geometries. When the C\sbond O bond length is fixed, four coordinates describe the geometry of the system: $R$, $\theta_{\text{A}}$, $\theta_{\text{B}}$, and $\phi$. The distance $R$ is the length of the vector $\bm{R}$ from the center of mass of monomer A to that of monomer B. The angles $\theta$ are between $\bm{R}$ and the vectors $\bm{r}_{\text{CO(A)}}$ and $\bm{r}_{\text{CO(B)}}$, which point from the C to the O atom in the respective monomer. The dihedral angle $\phi$ is between the planes spanned by ($\bm{R},\bm{r}_{\text{A}}$) and ($\bm{R},\bm{r}_{\text{B}}$). The 4D grid consists of 7 $\theta_{\text{A}}$ angles, 7 $\theta_{\text{B}}$ angles, 6 $\phi$ angles, and 13 $R$ values (3.5, 3.75, 4.0, 4.25, 4.5, 5.0, 5.5, 6.0, 6.5, 7.0, 9.0, 15.0 and 20.0~\AA). The angles were chosen in order to enable a spherical expansion of the interaction energy (see below). Calculations were performed for three C\sbond O distances, \emph{i.e.} molecule A was kept at the ground state equilibrium distance $r_e=1.128$~\AA\ while molecule B has $r_e$, $1.1r_e$, and $0.9r_e$. The interaction energy was found from CCSD(T) calculations using a standard aug-cc-pVQZ~\citep{Woon1993} basis set with the Boys-Bernardi counterpoise correction. All calculations were performed with the Molpro~\citep{MOLPRO} program.

A spherical expansion of the potential~\citep{VanHemert1983} was made to analyze the potential energy surface. This expansion was used to generate contour plots of the potential. When both molecules are in one plane, there are two minima separated by a barrier. The lowest minimum, with an interaction energy of -16.7~meV, occurs with the two~CO molecules parallel with $\theta_{\text{A}}$ and $\theta_{\text{B}}$ angles of 135$^{\circ}$ and -135$^{\circ}$ respectively, so with the two carbon atoms closest together.  The other minimum is at $-15.5$~meV and for $\theta_{\text{A}}$ and $\theta_{\text{B}}$ at 60$^\circ$ and -60$^\circ$ respectively, with now the oxygen atoms closest together. The center of mass distances for the minima are 4.5 and 3.7~\AA\ respectively.

For the AKMC simulations, the spherical expansion parametrization is too expensive. Instead, the \textit{ab-initio} interaction potential was parametrized as a site-site potential with electrostatic, exchange repulsion, dispersion, and intramolecular contributions:
\begin{equation}
V_{\text{CO\sbond CO}} = V_{\text{el-st}}+V_{\text{exch}}+V_{\text{disp}}+V_{\text{intra(A)}}+V_{\text{intra(B)}}   
\end{equation}
The electrostatic part, $V_{\text{el-st}}$, contains interactions between charges located on each atom and on the molecular centers of mass. The values of the charges are initially chosen to exactly reproduce the dipole and quadrupole moments at the specific $r_\text{CO}$. The moments were taken from MCSCF/CCI calculations with the aug-cc-pVQZ basis set. The exchange repulsion and dispersion terms are expressed as a Buckingham potential between the atomic sites in the dimer:
\begin{equation}
V_{\text{exch}} +V_{\text{disp}}=  \sum_{i\in\text{A}}\sum_{j\in\text{B}}  A_{ij} \exp{\left(-B_{ij}r_{ij}\right)}-\frac{C_{ij}}{r_{ij}^6}.
\end{equation}
The parameters $A_{ij}$ and $B_{ij}$ were optimized in a least squares procedure where all interaction energies with a value below 25~meV were included. The limit of 25~meV was chosen in order to focus the optimization of the potential on the bound part. The values of the $B_{ij}$ parameters were initially derived  from the relation between standard Lennard-Jones $C_{­6}$ and $C_{12}$ coefficients and the parameters used here, as given by~\cite{Lim2009}. The $B_{ij}$ parameters and the charges were then slightly adapted in order to minimize the least squares standard deviation. All parameters for the Buckingham potential are listed in Table~\ref{tab:table1}. It was found that the interaction energy could be well be represented only when the electrostatic term was made dependent on the intramolecular distance $r_\text{CO}$. More specifically, the changes in the charges were made proportional to the changes in computed charge as derived from the \textit{ab-initio} dipole and quadrupole moments:
\begin{equation}
q_i = q_i^0 \exp{\left(-\sigma_i(r_{\text{CO}} - r_e)\right)}.
\end{equation}
The charges $q_i^0$ are $-0.470$~$e$ on the C atom and $-0.615$~$e$ on the O atom. The values of $\sigma_i$ are $3.844$ and $2.132$~\AA$^{-1}$ for C and O, respectively. The parameters of the Buckingham potential remain independent of $r_\text{CO}$. The intramolecular interactions within each monomer are described with a Morse potential:
\begin{equation}
V_{\text{intra}} = D_e\left[1-\exp{\left(-\gamma(r_{\text{CO}} - r_e)\right)} \right]^2,
\end{equation}
where $D_e=11.23$~eV and $r_e=1.128$~\AA\ are the experimental dissociation energy and equilibrium bond length. These agree to within 0.1\% with the \textit{ab-initio} calculations. The $\gamma$ parameter was fitted to reproduce the \textit{ab-initio} bond length dependence of the potential energy and has a value of $2.328$~\AA$^{-1}$.

\subsubsection{Quality Of The Potential}
Contour plots were also generated from the site-site potential. Given the simple form of the parametrized potential, the agreement between the full \textit{ab-initio} contours and the model contours is satisfactory. In the in-plane case there are again two minimum energy structures separated by a barrier, the energy ordering of the minima is however reversed and the center of mass distances are somewhat different. The model molecules are softer than the \textit{ab-initio} molecules. This is in part due to the bias in the selection of configurations used in the least squares fit. Nevertheless, the use of these model parameters in molecular dynamics (MD) calculations of crystalline CO does not lead to large structural deviations from experiments, as becomes clear from the analysis below.

The quality of the potential was tested by a series of MD simulations. These were based on the analytical forces derived from the site-site potential, the velocity Verlet integration scheme, and a Berendsen thermostat~\citep{AllenTildesley}.
As a first test, pure amorphous CO crystals consisting of 200, 300, 500, 800, and 1200 CO molecules were created. These crystals were grown by adding CO molecules, one by one, to the previous molecules. Each new molecule was positioned at $10$~\AA\ from the surface of the core formed by the molecules already present. The four angles describing the initial orientation and center of mass position of the new molecule with respect to the ones already present were determined by a random number generator. The energetically most favorable final position of the new molecule was then determined with the simplex method. In this procedure the positions of the molecules already present were kept frozen. The energies in the simplex procedure were derived from the site-site potential. When all CO molecules were added the crystal was made to undergo temperature cycling in an MD procedure, in 6~K steps from 0 to 30~K and back to 6~K. At each temperature the crystal was kept for 200000~a.u. of time ($\sim$5~ps). As a second test, a series of crystalline samples was made, with a structure close to that of $\alpha$-CO (P$2_13$). First an $8\times8\times8$ unit cell crystal was constructed using the standard crystal data as input~\citep{Vegard1930}. Then spherical cuts where made containing 221, 522, and 1055 CO molecules. Also these crystals underwent the temperature cycling procedure described above.

\begin{figure}[h]
\includegraphics{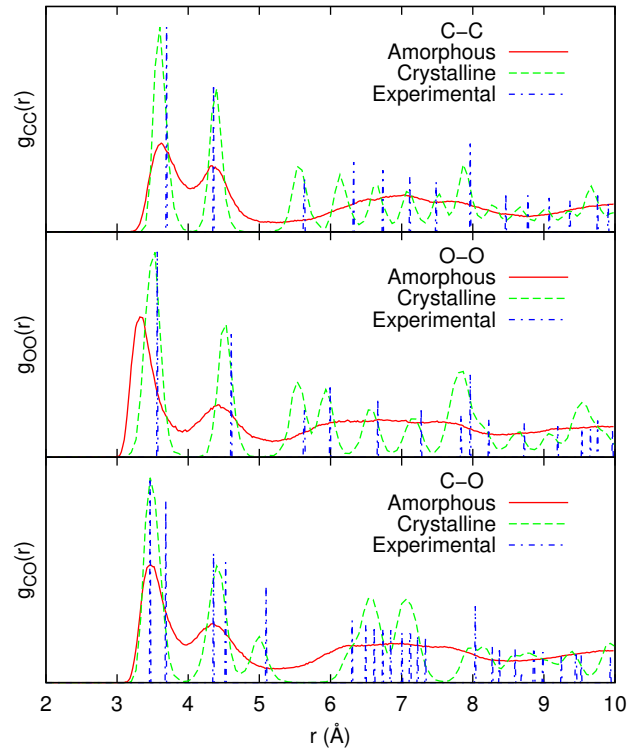}
\caption{Radial distribution functions, $g(r)$, in arbitrary units, from $6$~K molecular dynamics simulations of amorphous CO and crystalline $\alpha$-CO. Data from the experimental crystal structure of $\alpha$-CO~\citep{Vegard1930} is shown for reference\label{fig:RDF_CO}.}
\end{figure}

The structure of the various samples is clearest when looking at the radial distribution functions. Because there is little difference in the radial distribution functions for the various sizes we consider here only two specific examples: the amorphous sample containing 1200 molecules and the crystalline sample with 1055 molecules. The distribution functions for these samples are shown in Figure~\ref{fig:RDF_CO}, together with the distribution from the standard $\alpha$-CO crystal structure.
The distributions were obtained by averaging over the structures during the last 100 steps in the 5~ps re-equilibration run at $6$~K.  The broad distributions of the amorphous samples show clearly that these systems are still amorphous after the temperature cycling. The crystalline systems show much sharper peaks and also reflect a slightly higher density. The most notable difference between the two is the relatively short nearest neighbor O-O distance in the amorphous crystals when compared with the crystalline sample. Due to temperature broadening the split nearest neighbor peak seen in the experimental C-O distribution cannot be seen in the MD simulations. Energy minimizations of the $\alpha$-CO structure do show this feature, however the splitting is less pronounced. From the crystalline sample, we calculated the density at 6~K as 1.05~g~cm$^{-3}$. This compares well with the density of 1.03~g~cm$^{-3}$ derived from the X-ray based fcc unit cell length of 5.64~\AA\ from \cite{Krupskii1973}. The density of the amorphous systems is about 2\% lower than that of the crystalline samples. 

From the MD data, we have derived the specific heat of the crystal and made an estimate of the binding energy that would be obtained for an infinite crystal. The average specific heat corresponds to $5\times 10^{-4}$~eV~K$^{-1}$. By extrapolating the data to $0$~K we estimate the cohesive energies to be $81$~meV for the amorphous system and $84$~meV for the crystalline one. The latter compares well to the experimental value of $86$~meV for crystalline CO~\citep{Kelley1935}.

\section{ZERO POINT ENERGY CORRECTIONS}
\label{app:zpe}
The potential energy functions for the H$_2$O\sbond CO and CO\sbond CO interactions used in this paper were fitted directly to \textit{ab-initio} calculations. When a potential is fitted in this manner however, the zero point energy (ZPE) will not be included in the binding energies. For these kinds of system, the ZPE contribution can be quite significant though. To correctly include this contribution to the binding energies, we have estimated the magnitude of the ZPE in the H$_2$O\sbond CO and CO\sbond CO interactions and corrected all binding energies accordingly. For the semi-empirical TIP4P/2005f potential, this correction was already made implicitly when fitting the potential.


The contribution from the ZPE is straightforward to calculate in the harmonic approximation but this often gives unreliable results. Therefore, we first calculated by how much the harmonic calculations are in error with respect to accurate bound state calculations of the ground state energy in our fitted potentials. This was done for the gas phase H$_2$O\sbond CO complex and CO\sbond CO dimer. We will explain the procedure for the H$_2$O\sbond CO interactions below and then briefly give the results for CO\sbond CO interactions.

The method we used to calculate the ground state level of the H$_2$O\sbond CO complex is described in detail in~\cite{Groenenboom2000}. The theory is given in~\cite{Avoird1994}. In summary, the potential is accurately expanded as a function of the center of mass distance and the four angles needed to describe the geometry, at a fixed monomer geometries\footnote{This rigid molecule simplification can be made because the monomer frequencies barely shift due to the interaction with the other molecule.}. The expansion is made for the angular dependence at each point on a radial grid consisting of 113 equidistant points between 2.22 and 10.05 \AA. The ground state energy level of the complex described with this expanded potential was then obtained by diagonalizing the appropriate Hamiltonian. This Hamiltonian describes the internal rotations in the complex and the stretching of the intermolecular bond. The radial basis used for this calculation consisted of the five lowest energy eigenfunctions of the radial part of the Hamiltonian with the interactions calculated from the full potential at the angular geometry corresponding to the global minimum. Coupled internal rotor functions were used as the angular basis and the calculation was performed at zero total angular momentum, $J=0$. The dissociation energy, $D_0$, of the complex was found to be 21.3 meV, whereas the interaction energy in the global minimum, $D_e$, is 44.0 meV. This corresponds to a ZPE of 22.7 meV. The harmonic analysis of the vibrational frequencies of the original potential for the gas-phase complex gives a ZPE correction of 30.6 meV. An overestimate of 7.9~meV, more than 30\% with respect to the rovibrational calculations. 

In the harmonic approximation we also calculated the ZPE contribution from the H$_2$O\sbond CO interactions for every state we found from the AKMC simulations of Section~\ref{sec:binding} on the unconstrained amorphous ice substrates. The average contribution amounts to $19.0\pm4.8$~meV, considerably less than the contribution to the gas-phase complex. 

From these calculations we estimate that the ZPE correction of $11.1\pm4.8$ meV for systems with a CO molecule in a water-rich environment. All binding energies presented in this paper have been corrected by this amount. In our previous work on CO dynamics on hexagonal ice~\citep{Karssemeijer2012} we did not consider this effect and we have to assume that binding energies presented there are also too strong by about 11.1 meV, although a slightly different H$_2$O\sbond CO potential was used there.

In the same way as for the H$_2$O\sbond CO potential, we made an estimate of the ZPE correction on the CO\sbond CO interactions for a CO molecule adsorbed on a CO-dominated ice. In this work, this correction only plays a role in Section~\ref{sec:highercoverages}, where binding energies of CO on a CO surface (at the higher coverages) are calculated. The correction was found to be $14.2\pm3.0$~meV and the binding energies we corrected accordingly.

\bibliographystyle{apj}
\bibliography{karssemeijer_2711}

\begin{thebibliography}{}
\expandafter\ifx\csname natexlab\endcsname\relax\def\natexlab#1{#1}\fi

\bibitem[{Abascal \& Vega(2005)}]{Abascal2005}
Abascal, J. L.~F., \& Vega, C. 2005, J. Chem. Phys., 123, 234505

\bibitem[{Al-Halabi {et~al.}(2004{\natexlab{a}})Al-Halabi, Fraser, Kroes, \&
  van Dishoeck}]{Al-Halabi2004a}
Al-Halabi, A., Fraser, H.~J., Kroes, G.~J., \& van Dishoeck, E.~F.
  2004{\natexlab{a}}, A\&A, 422, 777

\bibitem[{Al-Halabi {et~al.}(2004{\natexlab{b}})Al-Halabi, van Dishoeck, \&
  Kroes}]{Al-Halabi2004}
Al-Halabi, A., van Dishoeck, E.~F., \& Kroes, G.~J. 2004{\natexlab{b}}, J.
  Chem. Phys., 120, 3358

\bibitem[{Allamandola {et~al.}(1999)Allamandola, Bernstein, Sandford, \&
  Walker}]{Allamandola1999}
Allamandola, L.~J., Bernstein, M.~P., Sandford, S.~A., \& Walker, R.~L. 1999,
  Space Sci. Rev., 90, 219

\bibitem[{Allen \& Tildesley(1989)}]{AllenTildesley}
Allen, M.~P., \& Tildesley, D.~J. 1989, {Computer Simulation of Liquids}
  (Oxford University Press, USA)

\bibitem[{Allodi {et~al.}(2013)Allodi, Ioppolo, Kelley, McGuire, \&
  Blake}]{Allodi2013}
Allodi, M.~A., Ioppolo, S., Kelley, M.~J., McGuire, B.~A., \& Blake, G.~A.
  2013, Phys. Chem. Chem. Phys., submitted

\bibitem[{Allouche {et~al.}(1998)Allouche, Verlaque, \& Pourcin}]{Allouche1998}
Allouche, A., Verlaque, P., \& Pourcin, J. 1998, J. Phys. Chem. B., 5647, 89

\bibitem[{Ayotte {et~al.}(2001)Ayotte, Smith, Stevenson, Dohn\'{a}lek, Kimmel,
  \& Kay}]{Ayotte2001}
Ayotte, P., Smith, R.~S., Stevenson, K.~P., {et~al.} 2001, J. Geophys. Res.,
  106, 387

\bibitem[{Bar-Nun {et~al.}(1985)Bar-Nun, Herman, Laufer, \&
  Rappaport}]{Bar-nun1985}
Bar-Nun, A., Herman, G., Laufer, D., \& Rappaport, M.~L. 1985, Icarus, 63, 317

\bibitem[{Batista \& J\'{o}nsson(2001)}]{Batista2001}
Batista, E.~R., \& J\'{o}nsson, H. 2001, Comp. Mater. Sci., 20, 325

\bibitem[{Bortz {et~al.}(1975)Bortz, Kalos, \& Lebowitz}]{Bortz1975}
Bortz, A.~B., Kalos, M.~H., \& Lebowitz, J.~L. 1975, J. Comput. Phys., 18, 10

\bibitem[{Bossa {et~al.}(2012)Bossa, Isokoski, de~Valois, \&
  Linnartz}]{Bossa2012}
Bossa, J.~B., Isokoski, K., de~Valois, M.~S., \& Linnartz, H. 2012, A\&A, 82, 1

\bibitem[{Bouwman {et~al.}(2007)Bouwman, Ludwig, Awad, \"{O}berg, Fuchs, van
  Dishoeck, \& Linnartz}]{Bouwman2007}
Bouwman, J., Ludwig, W., Awad, Z., {et~al.} 2007, A\&A, 476, 995

\bibitem[{Chang {et~al.}(2005)Chang, Cuppen, \& Herbst}]{Chang2005}
Chang, Q., Cuppen, H.~M., \& Herbst, E. 2005, A\&A, 611, 599

\bibitem[{Chang \& Herbst(2012)}]{Chang2012}
Chang, Q., \& Herbst, E. 2012, ApJ, 759, 147

\bibitem[{Charnley(1998)}]{Charnley1998}
Charnley, S.~B. 1998, ApJL, 509, L121

\bibitem[{Charnley {et~al.}(1992)Charnley, Tielens, \& Millar}]{Charnley1992}
Charnley, S.~B., Tielens, A. G. G.~M., \& Millar, T.~J. 1992, ApJL, 399, L71

\bibitem[{Collings {et~al.}(2003{\natexlab{a}})Collings, Dever, Fraser, \&
  McCoustra}]{Collings2003}
Collings, M.~P., Dever, J.~W., Fraser, H.~J., \& McCoustra, M. R.~S.
  2003{\natexlab{a}}, Ap\&SS, 285, 633

\bibitem[{Collings {et~al.}(2003{\natexlab{b}})Collings, Dever, Fraser,
  McCoustra, \& Williams}]{Collings2003b}
Collings, M.~P., Dever, J.~W., Fraser, H.~J., McCoustra, M. R.~S., \& Williams,
  D.~A. 2003{\natexlab{b}}, ApJ, 583, 1058

\bibitem[{Cuppen \& Garrod(2011)}]{Cuppen2011a}
Cuppen, H.~M., \& Garrod, R.~T. 2011, A\&A, 529, A151

\bibitem[{Cuppen \& Herbst(2005)}]{Cuppen2005}
Cuppen, H.~M., \& Herbst, E. 2005, MNRAS, 361, 565

\bibitem[{Cuppen {et~al.}(2011)Cuppen, Penteado, Isokoski, van~der Marel, \&
  Linnartz}]{Cuppen2011B}
Cuppen, H.~M., Penteado, E.~M., Isokoski, K., van~der Marel, N., \& Linnartz,
  H. 2011, MNRAS, 417, 2809

\bibitem[{Cuppen {et~al.}(2009)Cuppen, van Dishoeck, Herbst, \&
  Tielens}]{Cuppen:2009}
Cuppen, H.~M., van Dishoeck, E.~F., Herbst, E., \& Tielens, A. G. G.~M. 2009,
  A\&A, 508, 275

\bibitem[{Devlin(1992)}]{Devlin1992}
Devlin, J.~P. 1992, J. Phys. Chem., 96, 6185

\bibitem[{Dohn\'{a}lek {et~al.}(2003)Dohn\'{a}lek, Kimmel, Ayotte, Smith, \&
  Kay}]{Dohnalek2003}
Dohn\'{a}lek, Z., Kimmel, G.~a., Ayotte, P., Smith, R.~S., \& Kay, B.~D. 2003,
  J. Chem. Phys., 118, 364

\bibitem[{Fuchs {et~al.}(2009)Fuchs, Cuppen, Ioppolo, Romanzin, Bisschop,
  Andersson, van Dishoeck, \& Linnartz}]{Fuchs2009}
Fuchs, G.~W., Cuppen, H.~M., Ioppolo, S., {et~al.} 2009, A\&A, 505, 629

\bibitem[{Garrod {et~al.}(2008)Garrod, Weaver, \& Herbst}]{Garrod:2008b}
Garrod, R.~T., Weaver, S. L.~W., \& Herbst, E. 2008, ApJ, 682, 283

\bibitem[{Gerakines {et~al.}(1995)Gerakines, Schutte, Greenberg, \& van
  Dishoeck}]{Gerakines1995}
Gerakines, P.~A., Schutte, W.~A., Greenberg, J.~M., \& van Dishoeck, E.~F.
  1995, A\&A, 296, 810

\bibitem[{Gibb {et~al.}(2000)Gibb, Whittet, Schutte, Boogert, Chiar,
  Ehrenfreund, Gerakines, Keane, Tielens, van Dishoeck, Kerkhof, \&
  Al}]{Gibb2000}
Gibb, E.~L., Whittet, D. C.~B., Schutte, W.~A., {et~al.} 2000, ApJ, 347

\bibitem[{Gillespie(1976)}]{Gillespie1976}
Gillespie, D.~T. 1976, J. Comput. Phys., 22, 403

\bibitem[{Gonz\'{a}lez \& Abascal(2011)}]{Gonzalez2011}
Gonz\'{a}lez, M.~A., \& Abascal, J. L.~F. 2011, J. Chem. Phys., 135, 224516

\bibitem[{Groenenboom {et~al.}(2000)Groenenboom, Wormer, \& van~der
  Avoird}]{Groenenboom2000}
Groenenboom, G.~C., Wormer, P. E.~S., \& van~der Avoird, A. 2000, J. Chem.
  Phys., 113, 6702

\bibitem[{Hama {et~al.}(2012)Hama, Kuwahata, Watanabe, Kouchi, Kimura, Chigai,
  \& Pirronello}]{Hama2012}
Hama, T., Kuwahata, K., Watanabe, N., {et~al.} 2012, ApJ, 757, 185

\bibitem[{Hasegawa {et~al.}(1992)Hasegawa, Herbst, \& Leung}]{Hagasewa1992}
Hasegawa, T.~I., Herbst, E., \& Leung, C.~M. 1992, ApJS, 82, 167

\bibitem[{Henkelman \& J\'{o}nsson(1999)}]{Henkelman1999}
Henkelman, G., \& J\'{o}nsson, H. 1999, J. Chem. Phys., 111, 7010

\bibitem[{Henkelman \& J\'onsson(2001)}]{Henkelman2001}
Henkelman, G., \& J\'onsson, H. 2001, J. Chem. Phys., 115, 9657

\bibitem[{Herbst \& van Dishoeck(2009)}]{Herbst2009a}
Herbst, E., \& van Dishoeck, E.~F. 2009, ARA\&A, 47, 427

\bibitem[{Ioppolo {et~al.}(2011{\natexlab{a}})Ioppolo, Cuppen, van Dishoeck, \&
  Linnartz}]{Ioppolo2011a}
Ioppolo, S., Cuppen, H.~M., van Dishoeck, E.~F., \& Linnartz, H.
  2011{\natexlab{a}}, MNRAS, 410, 1089

\bibitem[{Ioppolo {et~al.}(2011{\natexlab{b}})Ioppolo, van Boheemen, Cuppen,
  van Dishoeck, \& Linnartz}]{Ioppolo2011}
Ioppolo, S., van Boheemen, Y., Cuppen, H.~M., van Dishoeck, E.~F., \& Linnartz,
  H. 2011{\natexlab{b}}, MNRAS, 413, 2281

\bibitem[{Jenniskens \& Blake(1994)}]{Jenniskens1994}
Jenniskens, P., \& Blake, D.~F. 1994, Science, 265, 753

\bibitem[{Karssemeijer {et~al.}(2012)Karssemeijer, Pedersen, J\'{o}nsson, \&
  Cuppen}]{Karssemeijer2012}
Karssemeijer, L.~J., Pedersen, A., J\'{o}nsson, H., \& Cuppen, H.~M. 2012,
  Phys. Chem. Chem. Phys., 14, 10844

\bibitem[{Kelley(1935)}]{Kelley1935}
Kelley, K.~K. 1935, U.S. Bur. Mines., 383, 34

\bibitem[{Kellogg(1994)}]{Kellogg1994}
Kellogg, G.~L. 1994, Surf. Sci. Rep., 21, 1

\bibitem[{Krupskii {et~al.}(1973)Krupskii, Prokhvatilov, Erenburg, \&
  Yantsevich}]{Krupskii1973}
Krupskii, I.~N., Prokhvatilov, A.~I., Erenburg, A.~I., \& Yantsevich, L.~D.
  1973, Phys. Status Solidi A, 19, 519

\bibitem[{Lim(2009)}]{Lim2009}
Lim, T.-c. 2009, Z. Naturforsch. A, 64, 200

\bibitem[{Loeffler {et~al.}(2005)Loeffler, Baratta, Palumbo, Strazzulla, \&
  Baragiola}]{Loeffler2005}
Loeffler, M.~J., Baratta, G., Palumbo, M.~E., Strazzulla, G., \& Baragiola,
  R.~A. 2005, A\&A, 435, 587

\bibitem[{Malek \& Mousseau(2000)}]{Malek2000}
Malek, R., \& Mousseau, N. 2000, Phys. Rev. E., 62, 7723

\bibitem[{Manca {et~al.}(2001)Manca, Martin, Allouche, \& Roubin}]{Manca2001}
Manca, C., Martin, C., Allouche, S., \& Roubin, P. 2001, J. Phys. Chem. B.,
  105, 12861

\bibitem[{Manca {et~al.}(2000)Manca, Roubin, \& Martin}]{Manca2000}
Manca, C., Roubin, P., \& Martin, C. 2000, Chem. Phys. Lett., 330, 21

\bibitem[{Matar {et~al.}(2008)Matar, Congiu, Dulieu, Momeni, \&
  Lemaire}]{Matar2008}
Matar, E., Congiu, E., Dulieu, F., Momeni, A., \& Lemaire, J.~L. 2008, A\&A,
  492, L17

\bibitem[{Minissale {et~al.}(2013)Minissale, Congiu, Baouche, Chaabouni,
  Moudens, Dulieu, Accolla, Cazaux, Manic\'{o}, \& Pirronello}]{Minissale2013}
Minissale, M., Congiu, E., Baouche, S., {et~al.} 2013, Phys. Rev. Lett., 111,
  053201

\bibitem[{Mispelaer {et~al.}(2013)Mispelaer, Theul\'{e}, Aouididi, Noble,
  Duvernay, Danger, Roubin, Morata, Hasegawa, \& Chiavassa}]{Mispelaer2013}
Mispelaer, F., Theul\'{e}, P., Aouididi, H., {et~al.} 2013, A\&A, A13, 1

\bibitem[{Noble {et~al.}(2012)Noble, Congiu, Dulieu, \& Fraser}]{Noble2012a}
Noble, J.~A., Congiu, E., Dulieu, F., \& Fraser, H.~J. 2012, MNRAS., 421, 768

\bibitem[{\"{O}berg {et~al.}(2011)\"{O}berg, Boogert, Pontoppidan, van~den
  Broek, van Dishoeck, Bottinelli, Blake, \& Evans}]{Oberg2011}
\"{O}berg, K.~I., Boogert, A. C.~A., Pontoppidan, K.~M., {et~al.} 2011, ApJ,
  740, 109

\bibitem[{Olsen {et~al.}(2004)Olsen, Kroes, Henkelman, Arnaldsson, \&
  J\'{o}nsson}]{Olsen2004}
Olsen, R.~A., Kroes, G.~J., Henkelman, G., Arnaldsson, A., \& J\'{o}nsson, H.
  2004, J. Chem. Phys., 121, 9776

\bibitem[{Pedersen {et~al.}(2012)Pedersen, Berthet, \&
  J\'{o}nsson}]{Pedersen2012a}
Pedersen, A., Berthet, J.~C., \& J\'{o}nsson, H. 2012, Lect. Notes. Comput.
  Sc., 7134, 34

\bibitem[{Pedersen \& J\'{o}nsson(2010)}]{Pedersen2010}
Pedersen, A., \& J\'{o}nsson, H. 2010, Math. Comput. Simulat., 80, 1487

\bibitem[{Perets {et~al.}(2005)Perets, Biham, Manico, Pirronello, Roser,
  Swords, \& Vidali}]{Perets2005}
Perets, H.~B., Biham, O., Manico, G., {et~al.} 2005, ApJ, 627, 850

\bibitem[{Pontoppidan(2006)}]{Pontoppidan2006}
Pontoppidan, K.~M. 2006, A\&A, 50, L47

\bibitem[{Pontoppidan {et~al.}(2005)Pontoppidan, Fraser, Dartois, Thi, van
  Dishoeck, Boogert, D'Hendecourt, Tielens, \& Bisschop}]{Pontoppidan2005}
Pontoppidan, K.~M., Fraser, H.~J., Dartois, E., {et~al.} 2005, A\&A, 1007, 981

\bibitem[{Ruffle \& Herbst(2002)}]{Ruffle2002}
Ruffle, D.~P., \& Herbst, E. 2002, MNRAS, 319, 837

\bibitem[{Sandford {et~al.}(1988)Sandford, Allamandola, Tielens, \&
  Valero}]{Sandford1988}
Sandford, S.~A., Allamandola, L.~J., Tielens, A. G. G.~M., \& Valero, G.~J.
  1988, ApJ, 329, 498

\bibitem[{Stevenson {et~al.}(1999)Stevenson, Kimmel, Dohn\'{a}lek, {Scott
  Smith}, \& Kay}]{Stevenson1999}
Stevenson, K.~P., Kimmel, G.~a., Dohn\'{a}lek, Z., {Scott Smith}, R., \& Kay,
  B.~D. 1999, Science, 283, 1505

\bibitem[{van~der Avoird {et~al.}(1994)van~der Avoird, Wormer, \&
  Moszynski}]{Avoird1994}
van~der Avoird, A., Wormer, P. E.~S., \& Moszynski, R. 1994, Chem. Rev., 94,
  1931

\bibitem[{van Dishoeck \& Blake(1998)}]{VanDishoeck1998}
van Dishoeck, E.~F., \& Blake, G.~a. 1998, ARA\&A, 36, 317

\bibitem[{van Hemert(1983)}]{VanHemert1983}
van Hemert, M.~C. 1983, J. Chem. Phys., 78, 2345

\bibitem[{Vasyunin \& Herbst(2013)}]{Vasyunin2013}
Vasyunin, A.~I., \& Herbst, E. 2013, ApJ, 762, 86

\bibitem[{Vega \& Abascal(2011)}]{Vega2011}
Vega, C., \& Abascal, J. L.~F. 2011, Phys. Chem. Chem. Phys., 13, 19663

\bibitem[{Vegard(1930)}]{Vegard1930}
Vegard, L. 1930, Zeitschrift f\"ur Physik, 61, 185

\bibitem[{Vineyard(1957)}]{Vineyard1957}
Vineyard, H. 1957, J. Phys. Chem. Solids, 3, 121

\bibitem[{Vissers {et~al.}(2003)Vissers, Wormer, \& van~der
  Avoird}]{Vissers2003}
Vissers, G. W.~M., Wormer, P. E.~S., \& van~der Avoird, A. 2003, Phys. Chem.
  Chem. Phys., 5, 4767

\bibitem[{Watanabe \& Kouchi(2002)}]{Watanabe2002}
Watanabe, N., \& Kouchi, A. 2002, ApJL, 571, 173

\bibitem[{Werner {et~al.}(2008)Werner, Knowles, Knizia, Manby, {Sch\"{u}tz},
  {et~al.}}]{MOLPRO}
Werner, H.-J., Knowles, P.~J., Knizia, G., {et~al.} 2008, MOLPRO, version
  2008.1, a package of ab initio programs, see http://www.molpro.net

\bibitem[{Woon \& Dunning(1993)}]{Woon1993}
Woon, D.~E., \& Dunning, T.~H. 1993, J. Chem. Phys, 98, 1358

\end{thebibliography}
\end{document}